\newif\ifUsenix\Usenixtrue 
\newif\ifAnon\Anonfalse 
\ifUsenix
\documentclass[letterpaper,twocolumn,10pt]{article}
\fi
\usepackage{hyperref}
\ifUsenix
  \usepackage{usenix2019_v3}
  \usepackage{authblk}
\else
  \documentclass[conference]{IEEEtran}
\fi
\usepackage{listings}
\usepackage{xcolor}
\usepackage{tabularx}
\usepackage[frozencache]{minted}
\setminted{fontsize=\scriptsize}
\usepackage[font=footnotesize,labelfont=bf]{caption}
\usepackage[square,comma,numbers,sort&compress]{natbib}

\AtBeginEnvironment{Verbatim}{%
  }

\usepackage{adjustbox}
\usepackage{nameref}
\usepackage{graphicx}
\usepackage{subcaption}
\usepackage{xspace}
\usepackage{microtype}
\usepackage{url}
\usepackage{booktabs}
\usepackage{amssymb}
\usepackage{enumitem}
\usepackage{xurl}
\usepackage{amsthm}
\usepackage[capitalise]{cleveref}
\usepackage{fancyvrb}
\usepackage{hyphenat}
\usepackage{multirow}
\usepackage{cleveref}
\usepackage{pbox}
\usepackage{tikz}
\usetikzlibrary{external}
\usepackage{letltxmacro}
\reversemarginpar
\usepackage{todonotes}
\usepackage{marginnote}
\usepackage{tikz}

\LetLtxMacro{\oldtodo}{\todo}
\renewcommand{\todo}[2][]{\tikzexternaldisable\oldtodo[fancyline,size=\footnotesize,#1]{#2}\tikzexternalenable}
\renewcommand{\todo}[1]{\tikzexternaldisable\oldtodo[fancyline,size=\footnotesize]{#1}\tikzexternalenable}

\usepackage{titling}
\usepackage[compact,nonindentfirst]{titlesec}

\def\chk{\tikz\fill[scale=0.4](0,.35) -- (.25,0) -- (1,.7) -- (.25,.15)
-- cycle;}

\definecolor{notecolor}{rgb}{0.8,0,0} 

\newcommand{\Red}[1]{{\color{red} #1}}

\newcommand{\tocite}[1]{\Red{[??]}}
\newcommand{\ignore}[1]{}

\newcommand{\etal}{et~al.\xspace}
\newcommand{\ie}{i.e.,\xspace}
\newcommand{\eg}{e.g.,\xspace}

\newenvironment{CompactItemize}%
{\begin{list}{$\blacktriangleright$}%
  {\leftmargin=\parindent \itemsep=2pt \topsep=2pt
    \parsep=0pt \partopsep=0pt}}%
{\end{list}}

\newcommand*\halfcirc[1][1ex]{%
  \begin{tikzpicture}
  \draw[fill] (0,0)-- (90:#1) arc (90:270:#1) -- cycle ;
  \draw (0,0) circle (#1);
  \end{tikzpicture}}
\newcommand*\fullcirc[1][1ex]{\tikz\fill (0,0) circle (#1);}

\theoremstyle{definition}

\fvset{fontsize=\small, xleftmargin=15pt, numbersep=5pt}

\def\C++{
  C\kern-.1667em\raise.30ex\hbox{\smaller{++}}%
  \spacefactor1000
}

\definecolor{xxxcolor}{rgb}{0.8,0,0}

\newcommand{\asm}[1]{\mintinline[fontsize=\small]{asm}{#1}}
\newmintinline[asmescaped]{asm}{fontsize=\small,escapeinside=||}

\newcommand{\mono}[1]{$\texttt{\small #1}$}

\newcommand\wasm{Wasm\xspace}
\newcommand\sys{Swivel\xspace}
\newcommand\Intel{Intel\textsuperscript{\textregistered}\xspace}
\newcommand\attackOneUpper{Sandbox breakout\xspace}

\newcommand\attackOne{sandbox breakout\xspace}
\newcommand\attackTwoUpper{Sandbox poisoning\xspace}

\newcommand\attackTwo{sandbox poisoning\xspace}
\newcommand\attackThreeUpper{Host poisoning\xspace}

\newcommand\attackThree{host poisoning\xspace}

\newcommand\sysStrawman{Strawman\xspace}
\newcommand\sysDesignOne{\sys-SFI\xspace}
\newcommand\sysDesignTwo{\sys-CET\xspace}

\newcommand\loadfence{LoadLfence\xspace}

\newcommand\lfence{\mono{lfence}\xspace}
\newcommand\lfences{{\lfence}s\xspace}

\newcommand\cet{Intel\textsuperscript{\textregistered} CET\xspace}
\newcommand\mpk{Intel\textsuperscript{\textregistered} MPK\xspace}
\newcommand\us{$\mu$s\xspace}
\newcommand\mincut{Mincut\xspace}

\newcommand\sgLoadfenceOverhead{8.7$\times$\xspace}
\newcommand\sgStrawmanOverhead{6.9$\times$\xspace}
\newcommand\sgMincutOverhead{2.4$\times$\xspace}
\newcommand\sgSfiASLROverhead{5.5\%\xspace}
\newcommand\sgCetASLROverhead{4.2\%\xspace}
\newcommand\sgSfiFullOverhead{61.9\%\xspace}
\newcommand\sgCetFullOverhead{99.7\%\xspace}
\newcommand\sgPHTtoBTBOverhead{52.9\%\xspace}
\newcommand\sgInterlockOverhead{93.2\%\xspace}
\newcommand\sgStockUnrollOverhead{0.0\%\xspace}

\newcommand\specLoadfenceOverhead{12.5$\times$\xspace}
\newcommand\specStrawmanOverhead{4.3$\times$\xspace}
\newcommand\specMincutOverhead{2.8$\times$\xspace}
\newcommand\specSfiASLROverhead{3.4\%\xspace}
\newcommand\specCetASLROverhead{2.6\%\xspace}
\newcommand\specSfiFullOverhead{47.3\%\xspace}
\newcommand\specCetFullOverhead{96.3\%\xspace}
\newcommand\specPHTtoBTBOverhead{38.5\%\xspace}
\newcommand\specInterlockOverhead{53.4\%\xspace}
\newcommand\specLoadfenceMinOverhead{7.3$\times$\xspace}
\newcommand\specLoadfenceMaxOverhead{19.6$\times$\xspace}
\newcommand\specStrawmanMinOverhead{1.8$\times$\xspace}
\newcommand\specStrawmanMaxOverhead{7.3$\times$\xspace}
\newcommand\specMincutMinOverhead{1.2$\times$\xspace}
\newcommand\specMincutMaxOverhead{5.4$\times$\xspace}
\newcommand\specSfiASLRMaxOverhead{10.3\%\xspace}
\newcommand\specCetASLRMaxOverhead{6.1\%\xspace}
\newcommand\specASLRMaxOverhead{\specSfiASLRMaxOverhead}
\newcommand\specSfiFullMinOverhead{3.3\%\xspace}
\newcommand\specCetFullMinOverhead{8.0\%\xspace}
\newcommand\specSfiFullMaxOverhead{86.1\%\xspace}
\newcommand\specCetFullMaxOverhead{240.2\%\xspace}
\newcommand\specFullMinOverhead{3.3\%\xspace}
\newcommand\specFullMaxOverhead{240.2\%\xspace}
\newcommand\specStockUnrollOverhead{5.9\%\xspace}

\usepackage{mdframed}
\newmdenv[
leftmargin = 0pt,
innerleftmargin = 1em,
innertopmargin = 0pt,
innerbottommargin = 0pt,
innerrightmargin = 0pt,
rightmargin = 0pt,
linewidth = 3.5pt,
linecolor= gray!50,
topline = false,
rightline = false,
bottomline = false
]{myleftbar}

\newcommand\stockTransition{2.14\us}
\newcommand\stockTransitionCb{0.07\us}
\newcommand\cetTransition{2.29\us}
\newcommand\cetTransitionCb{0.08\us}
\newcommand\sfiTransition{4.5\us}
\newcommand\sfiTransitionCb{1.26\us}
\newcommand\cetTransitionAslr{4.08\us}
\newcommand\cetTransitionAslrCb{0.79\us}

\newcommand\cdnMinOverhead{28.4\%\xspace}
\newcommand\cdnMaxOverhead{33.7\%\xspace}

\newcommand\cdnHash{Check SHA-256\xspace}
\newcommand\cdnJpgQuality{Change JPEG quality\xspace}
\newcommand\cdnTemplatedHTML{Templated HTML\xspace}
\newcommand\cdnXMLtoJSON{XML to JSON\xspace}
\newcommand\cdnML{Image classification\xspace}

\title{\Large
\textbf{Swivel: Hardening WebAssembly against Spectre}
}

\newcommand{\ucsdmark}{{$^{\dagger}$}\xspace}
\newcommand{\wpimark}{{$^{\mathparagraph}$}\xspace}
\newcommand{\intelmark}{{$^{\ast}$}\xspace}
\newcommand{\utmark}{{$^{\ddagger}$}\xspace}
\newcommand{\intellabsmark}{{$^{\star}$}\xspace}

\ifAnon
  \author{}
\else
  \author{
    Shravan Narayan\ucsdmark
    \quad
    Craig Disselkoen\ucsdmark
    \quad
    Daniel Moghimi\wpimark\ucsdmark
    \\
    Sunjay Cauligi\ucsdmark
    \quad
    Evan Johnson\ucsdmark
    \quad
    Zhao Gang\ucsdmark
    \\
    Anjo Vahldiek-Oberwagner\intellabsmark
    \quad
    Ravi Sahita\intelmark
    \quad
    Hovav Shacham\utmark
    \quad
    Dean Tullsen\ucsdmark
    \quad
    Deian Stefan\ucsdmark
    \\
    \small{
\ucsdmark{UC San Diego}\qquad
\wpimark{Worcester Polytechnic Institute}\qquad
\intellabsmark{Intel Labs}\qquad
\intelmark{Intel}\qquad
\utmark{UT Austin}\qquad
}
  }
\fi

\date{\vspace{-1em}}
\hypersetup{breaklinks=true}
\begin{document}
\setlength{\droptitle}{-2em}

\maketitle
\thispagestyle{plain}
  \frenchspacing
\pagestyle{plain}

\begin{abstract}
We describe Swivel, a new compiler framework for hardening WebAssembly (Wasm)
  against Spectre attacks.
Outside the browser, Wasm has become a popular lightweight, in-process sandbox
  and is, for example, used in production to isolate different clients on edge
  clouds and function-as-a-service platforms.
Unfortunately, Spectre attacks can bypass Wasm's isolation guarantees.
Swivel hardens Wasm against this class of attacks by ensuring that potentially
malicious code can neither use Spectre attacks to break out of the Wasm
sandbox nor coerce victim code---another Wasm client or the embedding
process---to leak secret data.

We describe two Swivel designs, a software-only approach that can be used on
  existing CPUs, and a hardware-assisted approach that uses extension available
  in \Intel 11th generation CPUs.
For both, we evaluate a randomized approach that mitigates Spectre and a
deterministic approach that eliminates Spectre altogether.
Our randomized implementations impose under \specASLRMaxOverhead overhead on the
Wasm-compatible subset of SPEC~2006, while our deterministic implementations
impose overheads between \specFullMinOverhead and \specFullMaxOverhead.
Though high on some benchmarks, Swivel's overhead is still between 9$\times$
  and 36.3$\times$ smaller than existing defenses that rely on pipeline fences.
\end{abstract}

\section{Introduction}
\label{sec:intro}
WebAssembly (\wasm) is a portable bytecode originally designed to safely run
native code (e.g., C/C++ and Rust) in the browser~\cite{wasm-paper}.
Since its initial design, though, \wasm has been increasingly used to
sandbox untrusted code outside the browser.
For example, Fastly and Cloudflare use \wasm to sandbox client applications
running on their edge clouds---where multiple client applications run within a
single process~\cite{fastly-wasm,cloudflare-workers}.
Mozilla uses \wasm to sandbox third-party C/C++ libraries in
Firefox~\cite{rlbox,rlbox-blog}.
Yet others use \wasm to isolate untrusted code in serverless
computing~\cite{hall2019execution}, IoT applications~\cite{microwasm},
games~\cite{flightsim}, trusted execution environments~\cite{enarxsgxwasm},
and even OS kernels~\cite{nebulet}.

In this paper, we focus on hardening Wasm against Spectre attacks---the class
of transient execution attacks which exploit control flow
predictors~\cite{Kocher2019spectre}.
Transient execution attacks which exploit features within the memory subsystem
(e.g., Meltdown~\cite{Lipp2018meltdown},
MDS~\cite{Schwarz2019ZL,Canella2019Fallout,VanSchaik2019RIDL}, and Load Value
Injection~\cite{vanbulck2020lvi}) are limited in scope and have already
been fixed in recent microarchitectures~\cite{intelMitigationList} (see
Section~\ref{sec:otherattacks}).
In contrast, Spectre can allow attackers to bypass \wasm's isolation boundary on
almost all superscalar
CPUs~\cite{AMDSpeculation,IntelSpeculation,AppleSpeculation}---and,
unfortunately, current mitigations for Spectre cannot be implemented entirely in
hardware~\cite{mcilroy2019spectre,yu2019speculative,barber2019specshield,koruyeh2019speccfi,venkman,RFC,jenkins2020ghostbusting,taram2019context}.

On multi-tenant serverless, edge-cloud, and function as a service (FaaS)
platforms, where Wasm is used as \emph{the} way to isolate mutually distursting
tenants, this is particulary concerning:\footnote{
Though our techniques are general, for simplicity we henceforth focus on Wasm
as used on FaaS platforms.
}
A malicious tenant can use Spectre to break out of the sandbox and read another
tenant's secrets in two steps (\S\ref{sec:pocs}).
First, they \emph{mistrain} different components of the underlying control flow
prediction---the conditional branch predictor (CBP), branch target buffer
(BTB), or return stack buffer (RSB)---to speculatively execute code that
accesses data outside the sandbox boundary.
Then, they \emph{reconstruct} the secret data from the underlying
microarchitectural state (typically the cache) using a side channel (cache
timing).
%

One way to mitigate such Spectre-based \emph{sandbox breakout attacks} is to
partition mutually distrusting code into separate processes.
By placing untrusted code in a separate process we can ensure that
the attacker cannot access secrets.
Chrome and Firefox, for example, do this by partitioning different \emph{sites}
into separate processes~\cite{chrome-site-isol, site-isolation-paper,
mozilla-sandbox}.
On a FaaS platform, we could similarly place tenants in separate processes.

Unfortunately, this would still leave tenants vulnerable to cross-process
\emph{\attackTwo
attacks}~\cite{IntelSpecMit,Canella2019,Koruyeh2018spectre5}.
Specifically, attackers can poison hardware predictors to coerce a victim
sandbox to speculatively execute gadgets that access secret data---from their
own memory region---and leak it via the cache (e.g., by branching on the
secret).
Moreover, using process isolation would sacrifice \wasm's scalability
(running many sandboxes within a process) and performance (cheap startup times
and context
switching)~\cite{fastly-wasm,cloudflare-workers,rlbox,hall2019execution}.

The other popular approach, removing speculation within the sandbox, is
also unsatisfactory.
For example, using pipeline fences to restrict \wasm code to sequential execution imposes
a \specStrawmanMinOverhead--\specStrawmanMaxOverhead slowdown on SPEC~2006
(\S\ref{sec:eval}).
Conservatively inserting pipeline fences before every dynamic load---an
approach inspired by the mitigation available in Microsoft's Visual Studio
compiler~\cite{mscvcspectre}---is even worse: it incurs
a \specLoadfenceMinOverhead--\specLoadfenceMaxOverhead overhead on SPEC
(\S\ref{sec:eval}).

In this paper, we take a compiler-based approach to hardening \wasm against
Spectre, without resorting to process isolation or the use of fences.
Our framework, \sys, addresses not only \attackOne and \attackTwo attacks, but
also \emph{\attackThree attacks}, i.e., Spectre attacks that coerce the process
hosting the \wasm sandboxes into leaking sensitive data.
That is, \sys ensures that a malicious \wasm tenant cannot
speculatively access data outside their sandbox nor coerce another tenant or
the host to divulge secrets of other sandboxes via poisoning.
We develop \sys via three contributions:

\paragraph{1. Software-only Spectre hardening (\S\ref{sec:sfi})}
Our first contribution, \sysDesignOne, is a software-only
approach to hardening \wasm against Spectre.
\sysDesignOne eliminates \attackOne attacks by compiling \wasm code to
\emph{linear blocks (LBs)}.
Linear blocks are straight-line x86 code blocks that satisfy two invariants:
(1) all transfers of control, including function calls, are at the block
boundary---to (and from) other linear blocks; and (2)
all memory accesses within a linear block are masked to the sandbox memory.
These invariants are necessary to ensure that the speculative control and data
flow of the \wasm code is restricted to the sandbox boundary.
They are not sufficient though: \sysDesignOne must also be tolerant of
possible RSB underflow.
We address this by (1) not emitting \mono{ret} instructions and therefore
completely bypassing the RSB and (2) using a \emph{separate stack} for return
addresses.

To address poisoning attacks, \sysDesignOne must still account for a poisoned
BTB or CBP.
Since these attacks are more sophisticated, we evaluate two different ways of
addressing them, and allow tenants to choose between them according to their
trust model.
The first approach uses address space layout randomization (ASLR) to randomize
the placement of each \wasm sandbox and flushes the BTB on each sandbox
boundary crossing.
This does not eliminate poisoning attacks; it only raises the bar of \wasm
isolation to that of process isolation.
Alternately, tenants can opt to eliminate these attacks altogether; to this
end, our deterministic \sysDesignOne rewrites conditional branches to indirect
jumps---thereby completely bypassing the CBP (which cannot be directly flushed)
and relying solely on the BTB (which can).

\paragraph{2. Hardware-assisted Spectre hardening (\S\ref{sec:cet})}
Our second contribution, \sysDesignTwo, restores the use of all predictors,
including the RSB and CBP, and partially obviates the need for BTB flushing.
It does this by sacrificing backwards compatibility and using new hardware security
extensions: Intel's Control-flow Enforcement Technology
(CET)~\cite{intel-manual} and Memory Protection Keys (MPK)~\cite{intel-manual}.

Like \sysDesignOne, \sysDesignTwo relies on linear blocks to address \attackOne
attacks.
But \sysDesignTwo does not avoid \mono{ret} instructions.
Instead, we use \cet's hardware \emph{shadow stack} to ensure that the RSB
cannot be misused to speculatively return to a location that is different from
the expected function return site on the stack~\cite{intel-manual}.

To eliminate \attackThree attacks, we use both \cet and \mpk.
In particular, we use \mpk to partition the application into two
domains---the host and \wasm sandbox(es)---and, on context switch,
ensure that each domain can only access its own memory regions.
We use \cet forward-edge control-flow integrity to ensure that application code
cannot jump, sequentially or speculatively, into arbitrary sandbox code (e.g.,
due to a poisoned BTB).
We do this by inserting \mono{endbranch} instructions in \wasm sandboxes to
demarcate valid jump targets, essentially partitioning the BTB into two
domains.
Our use of Intel's MPK and CET ensures that even if the host code
runs with poisoned predictors, it cannot read---and thus leak---sandbox data.

Since \mpk only supports 16 protection regions, we cannot use it to similarly prevent
\attackTwo attacks: serverless, edge-cloud, and FaaS platforms have thousands
of co-located tenants.
Hence, to address \attackTwo attacks, like for \sysDesignOne, we consider and evaluate
two designs.
The first (again) uses ASLR to randomize the location of each sandbox and
flushes the BTB on sandbox entry; we don't flush on sandbox exit since the
host can safely run with a poisoned BTB.
The second is deterministic and not only allows using conditional branches but
also avoids BTB flushes.
It does this using a new technique, \emph{register interlocking}, which tracks
the control flow of the \wasm sandbox and turns every misspeculated memory
access into an access to an empty guard page.
Register interlocking allows a tenant to run with a poisoned BTB or CBP
since any potentially poisoned speculative memory accesses will be invalidated.

\paragraph{3. Implementation and evaluation (\S\ref{sec:impl}--\ref{sec:eval})}
We implement both \sysDesignOne and \sysDesignTwo by modifying the Lucet
compiler's \wasm-to-x86 code generator (Cranelift) and runtime.
To evaluate \sys's security we implement proof of concept breakout and poisoning
attacks against stock Lucet (mitigated by \sys).
We do this for all three Spectre variants, i.e., Spectre attacks that abuse the
CBP, BTB, and RSB.

We evaluate \sys's performance against stock Lucet and several fence-insertion
techniques on several standard benchmarks.
On the \wasm compatible subset of the SPEC~2006 CPU benchmarking suite we find
the ASLR variants of \sysDesignOne and \sysDesignTwo impose little
overhead---they are at most \specSfiASLRMaxOverhead and \specCetASLRMaxOverhead
slower than Lucet, respectively.
Our deterministic implementations, which eliminate all three categories of
attacks, incur modest overheads: \sysDesignOne and \sysDesignTwo are respectively
\specSfiFullMinOverhead--\specSfiFullMaxOverhead (geomean:
\specSfiFullOverhead) and \specCetFullMinOverhead--\specCetFullMaxOverhead
(geomean: \specCetFullOverhead) slower than Lucet.
These overheads are smaller than the overhead imposed by state-of-the-art
fence-based techniques.

\paragraph{Open source and data}
We make all source and data available under an open source license at:
\url{https://swivel.programming.systems}.

\section{A brief overview of Wasm and Spectre}
\label{sec:background}
In this section, we give a brief overview of WebAssembly's use in multi-tenant
serverless and edge-cloud computing platforms---and more generally function as
a service (FaaS) platforms.
In particular, we describe how FaaS platforms use \wasm as an intermediate
compilation layer for isolating different tenants today.
We then briefly review Spectre attacks and describe how today's approach to
isolating \wasm code is vulnerable to this class of attacks.

\begin{figure}[t!]
  \centering
  \resizebox{.9\hsize}{!}{
    \includegraphics[width=1\hsize]{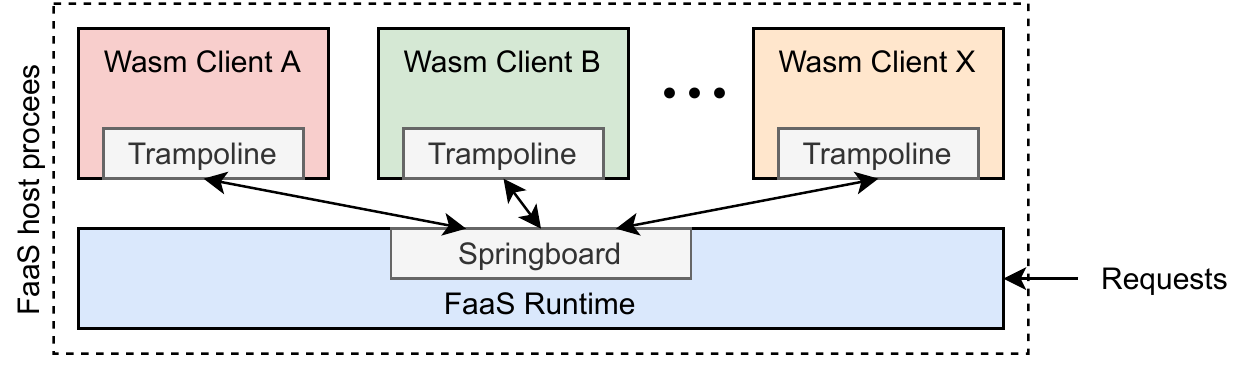}
  }
  \caption{FaaS platform using Wasm to isolate mutually distrusting tenants.}
  \label{fig:faas}
  \vspace{-.1in}
\end{figure}

\subsection{WebAssembly}
\label{subsec:wasm_bg}

\wasm is a low-level 32-bit machine language explicitly designed to embed C/C++
and Rust code in host applications.
\wasm programs (which are simply a collection of functions) are (1)
deterministic and well-typed, (2) follow a structured control flow discipline,
and (3) separate the heap---the \emph{linear memory}---from the well-typed
stack and program code.
These properties make it easy for compilers like Lucet to sandbox \wasm code and
safely embed it within an application~\cite{wasm-paper}.

\paragraph{Control flow safety}
Lucet's code generator, Cranelift,
ensures that the control flow of the compiled code is restricted
to the sandbox and cannot be bent to bypass bounds checks (e.g., via
return-oriented programming).
This comes directly from preserving \wasm's semantics during
compilation.
For example, compiled code preserves \wasm's safe stack~\cite{wasm-paper,
safestack}, ensuring that stack frames (and thus return values on the stack)
cannot be clobbered.
The compiled code also enforces \wasm's coarse-grained CFI and, for example,
matches the type of each indirect call site with the type of the target.

\paragraph{Memory isolation}
When Lucet creates a \wasm sandbox, it reserves 4GB of virtual memory for the
\wasm heap and uses Cranelift to bound all heap loads and stores.
To this end, Cranelift (1) explicitly passes a pointer to the base of the heap
as the first argument to each function and (2) masks all pointers to be within
this 4GB range.
Like previous software-based isolation (SFI) systems~\cite{wahbe-sfi},
Cranelift avoids expensive bounds check operations by using guard pages to trap
any offsets that may reach beyond the 4GB heap space.

\paragraph{Embedding \wasm}
An application with embedded \wasm code will
typically require context switching between the \wasm code and host---\eg to
read data from a socket.
Lucet allows safe control and data flow across the host-sandbox
boundary via \emph{springboards} and \emph{trampolines}.
Springboards are used to enter \wasm code---they set up the program context
for \wasm execution---while trampolines are used to restore the host context
and resume execution in the host.

\paragraph{Using \wasm in FaaS platforms}
WebAssembly FaaS platforms like Fastly's Terrarium allow clents to deploy
scalable function-oriented Web and cloud applications written in any langauge
(that can be compiled to Wasm).
Clients compile their code to \wasm and upload the resulting module to the
platform; the platform handles scaling and isolation.
As shown in Figure~\ref{fig:faas}, FaaS platforms place thousands of client
\wasm modules within a single host process, and distribute these processes
across thousands of servers and multiple datacenters.
This is the key to scaling---it allows any host process, in any
datacenter, to spawn a fresh \wasm sandbox instance and run any client function
in response to a user request.
By using \wasm as an intermediate layer, FaaS platforms isolate the client
for free~\cite{wahbe-sfi, fastly-wasm, gobi, wasm-il, lucet-talk, rlbox}.
Unfortunately, this isolation does not hold in the presence of Spectre attacks.

\subsection{Spectre attacks}
\label{subsec:spectre-attacks}

Spectre attacks exploit hardware predictors to induce mispredictions and
speculatively execute instructions---\emph{gadgets}---that would not run
sequentially~\cite{Kocher2019spectre}.
Spectre attacks are classified by the hardware predictor they exploit~\cite{Canella2019}.
We focus on the three Spectre variants that hijack control flow:

\begin{CompactItemize}
  \item{\bf Spectre-PHT}
    Spectre-PHT~\cite{Kocher2019spectre} exploits the \emph{pattern history table} (PHT),
    which is used as part of the \emph{conditional branch predictor} (CBP) to guess
    the direction of a conditional branch while the condition is still being evaluated.
    In a Spectre-PHT attack, the attacker (1) pollutes entries in the PHT so that a branch
    is mispredicted to the wrong path.
    The attacker can then use this wrong-path execution to
    bypass memory isolation guards or control flow integrity.

  \item{\bf Spectre-BTB}
    Spectre-BTB~\cite{Kocher2019spectre} exploits the \emph{branch target buffer} (BTB),
    which is used to predict the target of an indirect jump~\cite{branch-predictors}.
    In a Spectre-BTB attack, the attacker pollutes entries in the BTB, redirecting
    speculative control flow to an arbitrary target.
    Spectre-BTB can thus be used to speculatively execute gadgets that are
    not in the normal execution path (\eg to carry out a ROP-style attack).

  \item{\bf Spectre-RSB}
    Spectre-RSB~\cite{Maisuradze2018spectre5,Koruyeh2018spectre5} exploits the
    \emph{return stack buffer} (RSB), which memorizes the
    location of recently executed \asm{call} instructions to predict the targets
    of \asm{ret} instructions.
    In a Spectre-RSB attack, the attacker uses chains of \asm{call} or \asm{ret}
    instructions to over- or underflow the RSB, and redirect speculative control
    flow in turn.
\end{CompactItemize}
Spectre can be used \emph{in-place} or
\emph{out-of-place}~\cite{Canella2019}.
In an in-place attack, the attacker mistrains the prediction for a victim branch by
repeatedly executing the victim branch itself.
In an out-of-place attack, the attacker
finds a secondary branch that is \emph{congruent} to the victim branch---predictor
entries are indexed using a subset of address bits---and uses this
secondary branch to mistrain the prediction for the victim branch.

\subsection{Spectre attacks on FaaS platforms}
\label{sec:SpectreOnWasm}

A malicious FaaS platform client who can upload arbitrary \wasm code can force
the \wasm compiler to emit native code which is safe during sequential
execution, but uses Spectre to bypass \wasm's isolation guarantees
during speculative execution.
We identify three kinds of attacks (Figure~\ref{fig:wasm_spectre}):

\begin{figure}[t!]
  \centering
  \resizebox{.99\hsize}{!}{
    \includegraphics[width=1\hsize]{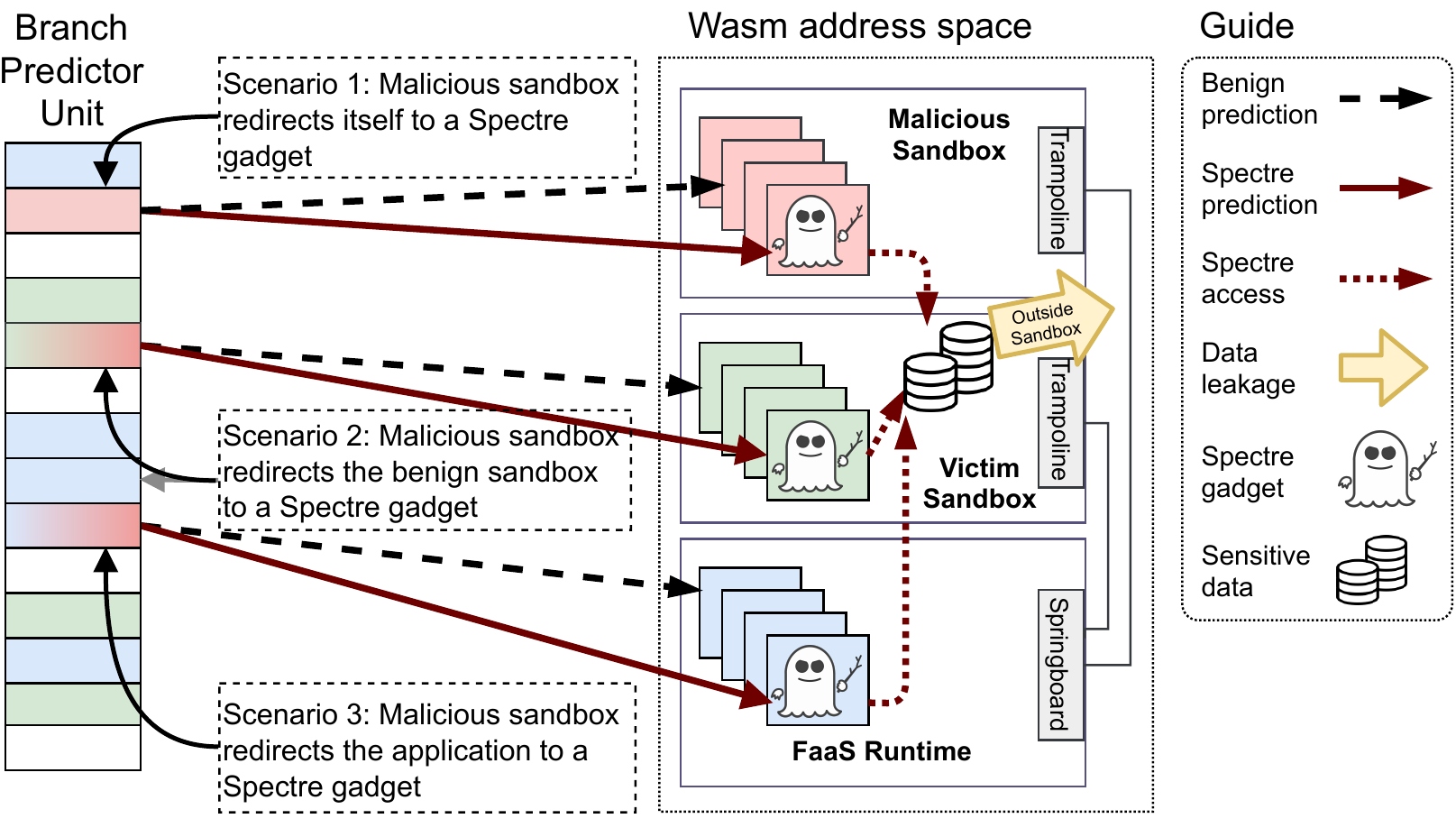}
  }
  \caption{A malicious tenant can fill branch predictors with invalid state
  (red).  In one scenario, the attacker causes its own branches to
  speculatively execute code that access memory outside of the sandbox. In the
  second and third scenarios, the attacker uses Spectre to respectively target
  a victim sandbox or the host runtime to misspeculate and leak secret data.
  }
  \label{fig:wasm_spectre}
\end{figure}

\begin{CompactItemize}
\item \textbf{Scenario 1: \attackOneUpper attacks}
The attacker bends the speculative control flow of their own
    module to access data outside the sandbox region.
For example, they can use Spectre-PHT to bypass conditional
bounds checks when accessing the indirect call table.
Alternatively, they can use Spectre-BTB to transfer the control flow into the
    middle of instructions to execute unsafe code (e.g., code that bypasses
    Wasm's implicit heap bounds checks).

\item \textbf{Scenario 2: \attackTwoUpper attacks}
The attacker uses an out-of-place Spectre attack to bend the control
flow of a victim sandbox and coerce the victim into leaking their own data.
Although this attack is considerably more sophisticated, we were still able to
implement proof of concept attacks following Canella et al.~\cite{Canella2019}.
Here, the attacker finds a (mispredicted) path in the victim sandbox that leads
to the victim leaking data, \eg through cache state.
They then force the victim to mispredict this path by using a congruent branch
within their own sandbox.

\item \textbf{Scenario 3: \attackThreeUpper attacks}
Instead of bending the control flow of the victim sandbox,
the attacker can use an out-of-place Spectre attack to bend the control flow of the
host runtime.
This
allows the attacker
    to speculatively access data from the host as well as any other sandbox.
\end{CompactItemize}

\begin{figure}

\begin{minted}{myasm_lexer.py:MyAsmLexer -x}
 mov    rdx,QWORD PTR [fn_table_len]  ; get fn table length
 cmp    rcx,rdx  ; check that rcx is in-bounds
 jb     index_ok
 ud2  ; trap otherwise
index_ok:
 lea    rdx,[fn_table]
 mov    rcx,QWORD PTR [rdx+rcx*4]
 call   rcx
\end{minted}
  \vspace{-.2in}
  \caption
  {
    A simplified snippet of the vulnerable code from our Spectre-PHT breakout attack.
    This code is safe during sequential execution (it checks the index
    \asmescaped{rcx} before using it to load a function table entry).
    But, during speculative execution, control flow may bypass this check and
    access memory outside the function table bounds.
  }
  \label{fig:attack_pinned_register}
\end{figure}

\noindent
\Cref{fig:attack_pinned_register} gives an example \attackOne gadget.
The gadget is in the implementation of the \wasm \asm{call_indirect}
instruction, which is used to call functions indexed in a module-level function
table.  This code first compares the function index \asmescaped{rcx} to the length
of the function table (to ensure that \asmescaped{rcx} points to a valid
entry). If \asmescaped{rcx} is valid, it then jumps to \asmescaped{index_ok}, loads
the function from the corresponding entry in the table, and calls it; otherwise
the code traps.

An attacker can mistrain the conditional branch on line 3 and cause it to
speculatively jump to \asmescaped{index_ok} even when \asmescaped{rcx} is
out-of-bounds.  By controlling the contents of \asmescaped{rcx}, the attacker
can thus execute arbitrary code locations outside the sandbox.
In Section~\ref{sec:pocs} we demonstrate several proof of concept attacks,
including a breakout attack that uses this gadget.
These attacks serve to highlight the importance of hardening \wasm against
Spectre.

\section{\sys: Hardening \wasm against Spectre}
\label{sec:design}

\begin{table}[tb!]
  \centering
  \footnotesize
  \begin{tabular}{p{1.5cm}p{1.5cm}|p{0.6cm}p{0.6cm}p{0.6cm}p{0.6cm}}

    \toprule

    \multirow{2}{4cm}{\textbf{\pbox{\textwidth}{Attack variant}}}
    &
    & \multicolumn{2}{c|}{\textbf{\pbox{\textwidth}{\sysDesignOne}}}
    & \multicolumn{2}{c}{\textbf{\pbox{\textwidth}{\sysDesignTwo}}}
    \\\cline{3-6}
    &
    & \multicolumn{1}{c|}{\textbf{\pbox{\textwidth}{ASLR}}}
    & \multicolumn{1}{c|}{\textbf{\pbox{\textwidth}{Det}}}
    & \multicolumn{1}{c|}{\textbf{\pbox{\textwidth}{ASLR}}}
    & \multicolumn{1}{c}{\textbf{\pbox{\textwidth}{Det}}} \\

    \midrule

    \multirow{2}{2cm}{Spectre-PHT}
    & in-place     & \fullcirc & \fullcirc & \fullcirc & \fullcirc \\
    & out-of-place & \halfcirc & \fullcirc & \halfcirc & \fullcirc \\

    \multirow{2}{2cm}{Spectre-BTB}
    & in-place     & \fullcirc & \fullcirc & \fullcirc & \fullcirc \\
    & out-of-place & \fullcirc & \fullcirc & \fullcirc & \fullcirc \\

    \multirow{2}{2cm}{Spectre-RSB}
    & in-place     & \fullcirc & \fullcirc & \fullcirc & \fullcirc \\
    & out-of-place & \fullcirc & \fullcirc & \fullcirc & \fullcirc \\

    \bottomrule

  \end{tabular}
  \caption{
    Effectiveness of \sys against different Spectre variants.
    A full circle indicates that \sys eliminates the attack while a half circle
    indicates that \sys only mitigates the attack.
  }
  \label{tab:swivel-effectiveness}
\end{table}

\sys extends Lucet---and the underlying Cranelift code generator---to address
Spectre attacks on FaaS \wasm platforms.
We designed \sys with several goals in mind.
First, \emph{performance}: \sys minimizes the number of pipeline fences it
inserts, allowing Wasm to benefit from speculative execution as much as
possible.
Second, \emph{automation}: \sys does not rely on user annotations or source
code changes to guide mitigations; we automatically apply mitigations when
compiling \wasm.
Finally, \emph{modularity}: \sys offers configurable protection, ranging from
probabilistic schemes with high performance to thorough mitigations with
strong guarantees.
This allows \sys users to choose the most appropriate mitigations (see
Table~\ref{tab:swivel-effectiveness}) according to their application domain,
security considerations, or particular hardware platform.

In the rest of this section, we describe our attacker model and
introduce a core abstraction: \emph{linear blocks}.
We then show how linear blocks, together with several other techniques, are
used to address both sandbox breakout and poisoning attacks.
These techniques span two \sys designs: \sysDesignOne, a software-only scheme which
provides mitigations compatible with existing CPUs; and \sysDesignTwo, which
uses hardware extensions (\cet and \mpk) available in the 11th generation \Intel
CPUs.

\paragraph{Attacker model}
\label{subsec:attacker_model}
We assume that the attacker is a FaaS platform client who can upload arbitrary
\wasm code which the platform will then compile and run alongside other clients
using \sys.
The goal of the attacker is to read data sensitive to another (victim) client
using Spectre attacks.
In the \sysDesignTwo case, we only focus on exfiltration via the data
cache---and thus assume an attacker who can only exploit gadgets that leak via
the data cache.
We consider transient attacks that exploit the memory subsystem (e.g.,
Meltdown~\cite{Lipp2018meltdown}, MDS~\cite{Schwarz2019ZL,Canella2019Fallout,VanSchaik2019RIDL},
and LVI~\cite{vanbulck2020lvi}) out of scope and discuss this in detail
in~\cref{sec:otherattacks}.

We assume that our Wasm compiler and runtime are correct.
We similarly assume the underlying operating system is secure and the
latest CPU microcode updates are applied.
We assume hyperthreading is disabled for any \sys scheme except for
the deterministic variant of \sysDesignTwo.
Consistent with previous findings~\cite{branch-predictors}, we assume BTBs predict 
the lower 32-bits of target addresses, while the upper 32-bits are inferred 
from the instruction pointer.

\sys addresses attackers that intentionally extract information using Spectre.
We do not prevent clients from accidentally leaking secrets during sequential
execution and, instead, assume they use techniques like constant-time
programming to prevent such leaks~\cite{cauligi2020foundations}.
For all \sys schemes except the deterministic variant of \sysDesignTwo, we
assume that a sandbox cannot directly invoke function calls in other
sandboxes, i.e., it cannot control the input to another sandbox to perform an
in-place poisoning attack.
We lastly assume that host secrets can be protected by placing them in a \wasm
sandbox, and discuss this further in \Cref{subsec:host-secrets}.

\medskip
\subsection{Linear blocks: local \wasm isolation}
\label{subsec:linear_blocks}

To enforce \wasm's isolation sequentially and speculatively, \sysDesignOne and
\sysDesignTwo compile \wasm code to linear blocks (LBs).
Linear blocks are straight-line code blocks that do not contain
\emph{any} control flow instructions except for their terminators---this
%
is in contrast to traditional basic blocks, which typically do not
consider function calls as terminators.
This simple distinction is important: It allows us to ensure that \emph{all} control
flow transfers---both sequential and speculative---land on linear block boundaries.
Then, by ensuring that individual linear blocks are \emph{safe}, we can ensure
that whole \wasm programs, when compiled, are confined and cannot violate
\wasm's isolation guarantees.

A linear block is safe if, independent of the control flow into the block,
\wasm's isolation guarantees are preserved.
In particular, we cannot rely on safety checks (\eg bounds checks for memory
accesses) performed across linear blocks since, speculatively, blocks may not
always execute in sequential order (\eg because of Spectre-BTB).
When generating native code, \sys ensures that a linear block is safe by:

\paragraph{Masking memory accesses}
Since we cannot make any assumptions about the initial contents of registers,
\sys ensures that unconditional heap bounds checks (performed via masking)
are performed in the same linear block as the heap memory access itself.
We do this by modifying the Cranelift optimization passes which could lift
bounds checks (e.g., loop invariant code motion) to ensure that they don't
move masks across linear block boundaries.
Similarly, since we cannot trust values on the stack, \sys ensures
that any value that is unspilled from the stack and used in a bounds
check is masked again.
We use this \emph{mask-after-unspill} technique to replace
Cranelift's unsafe mask-before-spill approach.
%

\paragraph{Pinning the heap registers}
To properly perform bounds checks for heap memory accesses, a \sys linear
block must determine the correct value of the heap base.
Unfortunately, as described above, we cannot make any assumptions about the
contents of any register or stack slot.
\sys thus reserves one register, which we call the \emph{pinned heap
register}, to store the address of the sandbox heap.
Furthermore, \sys prevents any
    instructions in the sandbox from altering the pinned heap register.
This allows each linear block to safely assume that the pinned heap register
holds the correct value of the heap base, even when the speculative control
flow of the program has gone awry due to
misprediction.

\paragraph{Hardening jump tables}
\wasm requires bounds checks on each access to indirect call tables and
switch tables.
\sys ensures that each of these bounds checks is local to the linear block
    where the access is performed. Moreover, \sys implements the bounds check
    using \emph{speculative load hardening}~\cite{RFC}, masking the index of
    the table access with the length of the table.
This efficiently prevents the attacker from speculatively bypassing the bounds check.

\sys does not check the indirect jump targets
beyond what Cranelift does.
At the language level, \wasm already guarantees that the targets of indirect jumps (\ie
the entries in indirect call tables and switch tables) can only be the tops of
functions or switch-case targets.
Compiled, these correspond to the start of \sys linear blocks.
Thus, an attacker can only train the BTB with entries that point to linear
blocks, which, by construction, are safe to execute in any context.

\paragraph{Protecting returns}
\wasm's execution stack---in particular, return addresses on the stack---cannot
be corrupted during sequential execution.
Unfortunately, this does not hold speculatively:
An attacker can write to the stack (\eg with a buffer overflow) and speculatively execute a return
instruction which will divert the control flow to their location
of choice.
\sys ensures that return addresses on the stack cannot be corrupted as such,
even speculatively.
We do this using a \emph{separate stack}
or \emph{shadow stack}~\cite{sok-shadowstacks}, as we detail below.


\begin{figure*}[t!]
  \centering
  \resizebox{.90\hsize}{!}{
    \includegraphics[width=1\hsize]{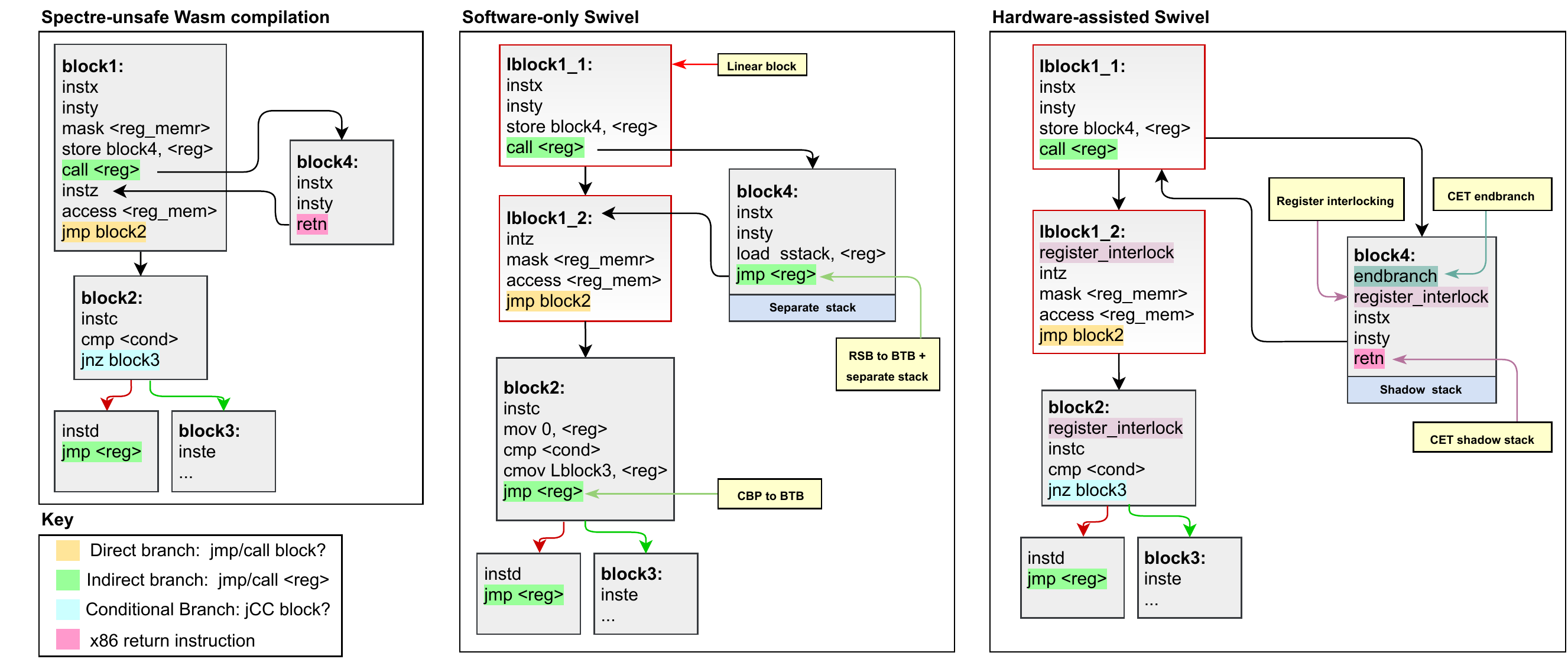}
  }
  \caption{\sys hardens \wasm against spectre via compiler transformations.
    In \sysDesignOne, we convert basic
    blocks to linear blocks. Each linear block (e.g., \mono{lblock1\_1}
    and
    \mono{lblock1\_2}) maintains local security guarantees for speculative
    execution.
    Then, we protect the backward edge (\mono{block4}) by replacing the return
    instructions
    and using a separate return stack. To eliminate poisoning attacks, in
    the deterministic version of \sysDesignOne, we further encode conditional
    branches as indirect jumps.
    In \sysDesignTwo, we similarly use linear blocks, but we allow return
    instruction,
    and protect returns using the hardware shadow stack. To reduce BTB
    flushes,
    we additionally use \cet's \mono{endbranch} to ensure that targets of
    indirect branches
    land at the beginning of linear blocks.
    In the deterministic version, we avoid BTB flushing and instead use register
    interlocking to prevent leakage on misspeculated paths.
  }
  \label{fig:design}
\end{figure*}

\subsection{\sysDesignOne}
\label{sec:sfi}

\sysDesignOne builds on top of linear blocks to address all three classes of
attacks.

\subsubsection{Addressing \attackOne attacks}

Compiling \wasm code to linear blocks eliminates most avenues for breaking out
of the sandbox.
The only way for an attacker to break out of the sandbox is to
speculatively jump into the middle of a linear block.
We prevent this with:

\paragraph{The separate stack}
We protect returns by preserving \wasm's safe return stack during
compilation.
Specifically, we create a separate stack in a fixed memory location, but outside
the sandbox stack and heap, to ensure that it cannot be overwritten by
sandboxed code.
We replace every call instruction with an instruction sequence that stores the
address of the subsequent instruction---the return address---to the next entry
in this separate stack. Similarly, we replace every return instruction with a
sequence that pops the address off the separate stack and jumps to that location.
To catch under- and over-flows, we surround the separate
stack with guard pages.

\paragraph{BTB flushing}
The other way an attacker can jump into the middle of a linear block is via a
mispredicted BTB entry.
Since all indirect jumps inside a sandbox can only point to the tops of linear
blocks, any such entries can only be set via a congruent entry outside any sandbox---\ie
an attacker must orchestrate the host runtime into mistraining a particular jump.
We prevent such attacks by flushing the BTB on transitions into and out
of the sandbox.\footnote{
In practice, BTB predictions are not absolute (as discussed in our attacker 
model), instead they are 32-bit offsets relative to the instruction 
pointer~\cite{branch-predictors}.
To ensure that this does not result in predictions at non linear block
boundaries, we restrict the sandbox code size to 4GB.
}

\subsubsection{Addressing sandbox and host poisoning attacks}

There are two ways for a malicious sandbox to carry out poisoning attacks:
By poisoning CBP or BTB entries.
Since we already flush the BTB to address \attackOne attacks, we trivially
prevent all BTB poisoning.
Addressing CBP poisoning is less straightforward.
We consider two schemes:

\paragraph{Mitigating CBP poisoning}
To mitigate CBP-based poisoning attacks, we use ASLR to randomize the layout of
sandbox code pages.
This mitigation is not sound---it is theoretically possible for an attacker
to influence a specific conditional branch outside of the sandbox.
As we discuss in Section~\ref{subsec:design_security},
this raises the bar to (at least) that of process isolation: The
attacker would would have to (1) de-randomize the ASLR of both their own
sandbox and the victim's and (2) find useful gadgets, which is itself an open problem (\S\ref{sec:related}).

\paragraph{Eliminating CBP poisoning}
\label{subsubsec:cbp-to-btb}
Clients that are willing to tolerate modest performance overheads
(\S\ref{sec:eval}) can opt to eliminate poisoning attacks.
We eliminate poisoning attacks by removing conditional branches from \wasm
sandboxes altogether.
Following~\cite{lee2017inferring}, we do this by using the \mono{cmov}
conditional move instruction to encode each conditional branch as an indirect
branch with only two targets (\Cref{fig:design}).

\subsection{\sysDesignTwo}
\label{sec:cet}

\sysDesignOne avoids using fences to address Spectre attacks, but ultimately
bypasses all but the BTB predictors---and even then we flush the BTB on every
sandbox transition.
\sysDesignTwo
uses \Intel CET~\cite{intel-manual} and \Intel MPK~\cite{intel-manual}
to restore the use of the CBP and RSB, and avoid BTB flushing.\footnote{
Appendix~\ref{app:cet} gives a brief introduction to these new hardware features.
}

\subsubsection{Addressing \attackOne attacks}

Like \sysDesignOne, we build on linear blocks to address \attackOne attacks
(\Cref{fig:design}).
\sysDesignTwo, however, prevents an attacker from speculatively jumping into
the middle of a linear block using:

\paragraph{The shadow stack}
\sysDesignTwo uses \cet's \emph{shadow stack} to protect returns.
Unlike \sysDesignOne's separate stack, the shadow stack is a
hardware-maintained stack, distinct from the ordinary data stack and
inaccessible via standard load and store instructions.
The shadow stack allows us to use call and return instructions as
usual---the CPU uses the shadow stack to check the integrity of return
addresses on the program stack during both sequential and speculative
execution.

\paragraph{Forward-edge CFI}
Instead of flushing the BTB, \sysDesignTwo uses \cet's coarse-grained control
flow integrity (CFI)~\cite{cfi} to ensure that sandbox code can only jump to
the top of a linear block.
We do this by placing an \mono{endbranch} instruction at the beginning of every
linear block that is used as an indirect target (\eg the start of a function
that is called indirectly).
During speculative execution, if the indirect branch predictor targets
an instruction other than an \mono{endbranch} (\eg inside the host runtime),
the CPU stops speculating~\cite{intel-manual}.

\paragraph{Conditional BTB flushing}
When using ASLR to address \attackTwo attacks, we still need to
flush the BTB on transitions into each sandbox.
Otherwise, one sandbox could potentially jump to a linear block in
another sandbox.
Our deterministic approach to \attackTwo (described below), however,
eliminates BTB flushes altogether.

\subsubsection{Addressing host poisoning attacks}
To prevent \attackThree attacks, \sysDesignTwo uses \mpk.
\mpk exposes new user mode instructions that allow a process to partition its
memory into sixteen linear regions and to selectively enable/disable read/write
access to any of those regions.
\sysDesignTwo uses only two of these protection domains---one for the host and one shared by
all sandboxes---and switches domains during the transitions between host and sandbox.
When the host creates a new sandbox, \sysDesignTwo allocates the heap memory
for the new sandbox with the sandbox protection domain, and then relinquishes
its own access to that memory so that it is no longer accessible by the host.
This prevents \attackThree attacks by ensuring that the host cannot be coerced
into leaking secrets from another sandbox.
We describe how we safely copy data across the
boundary later (\S\ref{sec:impl}).

\subsubsection{Addressing sandbox poisoning attacks}
\label{subsubsec:cet_sandbox_poison}
By poisoning CBP or BTB entries, a malicious sandbox can coerce a
victim sandbox into executing a gadget that leaks sensitive data.
As with \sysDesignOne, we consider both a probabilistic and deterministic
design to addressing these attacks.
Since the probabilistic approach is like \sysDesignOne's, we describe only the
deterministic design.

\paragraph{Preventing leaks under poisoned execution}
\sysDesignTwo does \emph{not} eliminate cross-sandbox CBP or BTB poisoning.
Instead, we ensure that a victim sandbox cannot be coerced into leaking
data via the cache when executing a mispredicted path.
To leak secrets through the cache, the attacker must maneuver the secret
data to a gadget that will use it as an offset into a memory region.
In Cranelift, any such gadget will use the heap, as stack memory
is always accessed at constant offsets from the stack pointer (which itself
cannot be directly assigned).
We thus need only prevent leaks that are via the Wasm heap---we
do this using \emph{register interlocks}.

\paragraph{Register interlocking}
\label{subsubsec:interlocks}
Our register interlocking technique tracks the control flow of a \wasm program
and prevents it from accessing its stack or heap when the speculative path
diverges from the sequential path.
We first assign each non-trivial linear block a unique 64-bit
\emph{block label}.
We then calculate the expected block label of every direct or indirect branch
and assign this value to a reserved \emph{interlock register} prior to branching.
At the beginning of each linear block, we check that the value of the
interlock register corresponds to the static block label using \mono{cmov}
instructions.
If the two do not match, we zero out the heap base register as well as the stack
register.
Finally, we unmap pages from the address space to ensure that any access from a
zero heap or stack base will fault---and thus will not affect cache state.

The register interlock fundamentally introduces a data dependency
between memory operations and the resolution of
control flow.
In doing so, we prevent any memory operations that would result in cache based
leaks, but do not prevent all speculative execution.
In particular, any arithmetic operations may still be executed speculatively.
This is similar to hardware taint tracking~\cite{yu2019speculative}, but
enforced purely through compiler changes.

Finally, \wasm also stores certain data (\eg globals variables and internal
structures) outside the \wasm stack or heap.
To secure these memory accesses with the register interlock, we introduce an
artificial heap load in the address computation for this data.

\subsection{Security and performance trade-offs}
\label{subsec:design_security}

\sys offers two design points for protecting \wasm modules from Spectre
attacks: \sysDesignOne and \sysDesignTwo.
For each of these schemes we further consider probabilistic (ASLR) and
deterministic techniques.
In this section, we discuss the performance and security trade-offs when
choosing between these various \sys schemes.

\begin{table}[t]
\footnotesize
\begin{tabular}{p{4cm}|p{0.6cm}p{0.6cm}|p{0.6cm}p{0.6cm}}

\toprule

\multirow{2}{4cm}{\textbf{\pbox{\textwidth}{\sys protection and technique}}}
& \multicolumn{2}{c|}{\textbf{\pbox{\textwidth}{\sysDesignOne}}}
& \multicolumn{2}{c}{\textbf{\pbox{\textwidth}{\sysDesignTwo}}} \\\cline{2-5}

& \multicolumn{1}{c|}{\textbf{\pbox{\textwidth}{ASLR}}}
& \multicolumn{1}{c|}{\textbf{\pbox{\textwidth}{Det}}}
& \multicolumn{1}{c|}{\textbf{\pbox{\textwidth}{ASLR}}}
& \multicolumn{1}{c}{\textbf{\pbox{\textwidth}{Det}}} \\

\midrule

\textbf{\attackOneUpper protections}   &      &      &      &      \\
- Linear blocks [CBP, BTB, RSB]        & \chk & \chk & \chk & \chk \\
- BTB flush in springboard [BTB]       & \chk & \chk &      &      \\
- Separate control stack [RSB]         & \chk & \chk &      &      \\
- CET endbranch [BTB]                  &      &      & \chk & \chk \\
- CET shadow stack [RSB]               &      &      & \chk & \chk \\
\midrule

\textbf{\attackTwoUpper protections}   &      &      &      &      \\
- BTB flush in springboard [BTB]       & \chk & \chk & \chk &      \\
- Code page ASLR [CBP]                 & \chk &      & \chk &      \\
- Direct branches to indirect [CBP]    &      & \chk &      &      \\
- Register interlock [CBP, BTB]        &      &      &      & \chk \\
\midrule

\textbf{\attackThreeUpper protections} &      &      &      &      \\
- Separate control stack [RSB]         & \chk & \chk &      &      \\
- Code page ASLR [CBP]                 & \chk &      &      &      \\
- BTB flush in trampoline [BTB]        & \chk & \chk &      &      \\
- Direct branches to indirect [CBP]    &      & \chk &      &      \\
- Two domain MPK [CBP]                 &      &      & \chk & \chk \\
\bottomrule

\end{tabular}
\caption{
Breakdown of \sys's individual protection techniques which, when combined,
address the thee different class of attacks on Wasm
(\S\ref{sec:SpectreOnWasm}).
For each technique we also list (in brackets) the underlying predictors.
}
\label{tab:mitigations}
\end{table}

\subsubsection{Probabilistic or deterministic?}

\Cref{tab:swivel-effectiveness} summarizes \sys's security guarantees.
\sys's deterministic schemes eliminate Spectre attacks, while
the probabilistic schemes eliminate Spectre attacks that exploit the BTB and
RSB, but trade-off security for performance when it comes to the CBP (\S\ref{sec:eval}):
Our probabilistic schemes only \emph{mitigate} Spectre attacks that exploit the
CBP.

To this end, (probabilistic) \sys hides branch offsets by randomizing code
pages. 
Previously, similar fine-grain approaches to address randomization have been
proposed to mitigate attacks based on return-oriented
programming~\cite{crane2015s,gens2017lazarus}.
Specifically, when loading a module, \sys copies the code pages of the \wasm
module to random destinations, randomizing all but
the four least significant bits (LSBs) to keep 16-byte alignment.
This method is more fine-grained than page remapping, which would fail to randomize the
lower 12 bits for 4KB instruction pages.

Unfortunately, only a subset of address bits are typically used by hardware
predictors.
Zhang et.  al~\cite{branch-predictors}, for example, found that only the 30
LSBs of the instruction address are used as input for BTB predictors.
Though a similar study has not been conducted for the CBP, if we
pessimistically assume that 30 LSBs are used for prediction then our
randomization offers at least 26 bits of entropy.
Since the attacker must de-randomize both their module and the victim
module, this is likely higher in practice.

As we show in Section~\ref{sec:eval}, the ASLR variants of \sys are faster than
the deterministic variants.
Using code page ASLR imposes less overhead than the deterministic techniques
(summarized in \Cref{tab:mitigations}).
This is not surprising: CBP conversion (in \sysDesignOne) and register
interlocking (in \sysDesignTwo) are the largest sources of performance
overhead.

For many application domains, this security-performance trade-off is
reasonable.
Our probabilistic schemes use ASLR only to mitigate \attackTwo
attacks---and
unlike \attackOne attacks, these attacks are significantly more challenging
for an attacker to carry out:
They must conduct an out-of-place attack on a specific target while accounting
for the unpredictable mapping of the branch predictor.
To our knowledge, such an attack has not been demonstrated, even
without the additional challenges of defeating ASLR.

Furthermore, on a FaaS platform, these attacks are even harder to pull off,
as the attacker has only a few hundred milliseconds to land an attack
on a victim sandbox instance before it finishes---and the
next victim instance will have entirely new mappings.
Previous work suggests that such an attack is not
practical in such a short time window~\cite{evtyushkin2018branchscope}.

For other application domains, the overhead of the deterministic \sys variants
may yet be reasonable. As we show in Section~\ref{sec:eval}, the average
(geometric) overhead of \sysDesignOne is \specSfiFullOverhead and that of
\sysDesignTwo is \specCetFullOverhead.
Moreover, users can choose to use \sysDesignOne and \sysDesignTwo according
to their trust model---\sys
allows sandboxes of both designs to coexist within a single process.

\subsubsection{Software-only or hardware-assisted?}

\sysDesignOne and \sysDesignTwo present two design points that have
different trade-offs beyond backwards compatibility.
We discuss their trade-offs, focusing on the deterministic variants.

\sysDesignOne eliminates Spectre attacks by allowing speculation only via the BTB predictor
and by controlling BTB entries through linear blocks and BTB flushing.
\sysDesignTwo, on the other hand, allows the other predictors.
To do this safely though, we use register interlocking  to create data
dependencies (and thus prevent speculation) on certain operations after
branches.
Our interlock implementation only guards \wasm memory operations---this means
that, unlike \sysDesignOne, \sysDesignTwo only prevents \emph{cache-based}
\attackTwo attacks.
While non-memory instructions (e.g., arithmetic operations) can still
speculatively execute, register interlocks sink performance: Indeed, the
overall performance overhead of \sysDesignTwo is higher than
\sysDesignOne~(\S\ref{sec:eval}).

At the same time, \sysDesignTwo can be used to handle a more powerful
attacker model (than our FaaS model).
First, \sysDesignTwo eliminates poisoning attacks even in the presence of
attacker-controlled input.
This is a direct corollary of being able to safely execute code with poisoned
predictors.
Second, \sysDesignTwo (in the deterministic scheme) is safe in the presence of
hyperthreading; our other \sys schemes assume that hyperthreading is disabled
(\S\ref{subsec:attacker_model}).
\sysDesignTwo allows hyperthreading because it doesn't rely on BTB flushing;
it uses register interlocking to eliminate sandbox poisoning attacks.
In contrast, our SFI schemes require the BTB to be isolated for the host and each \wasm sandbox---an
invariant that may not hold if, for example,
hyperthreading interleaves host application and \wasm code on sibling threads.

\section{Implementation}
\label{sec:impl}
We implement \sys on top of the Lucet \wasm compiler and
runtime~\cite{lucet-talk,fastly-wasm}.
In this section, we describe our modifications to Lucet.

We largely implement \sysDesignOne and \sysDesignTwo as passes in Lucet's code
generator Cranelift.
For both schemes, we add support for pinned heap registers and add direct
\mono{jmp} instructions to create linear block boundaries.
We modify Cranelift to harden switch-table and indirect-call accesses: Before
loading an entry from either table, we truncate the index to the length
of the table (or the next power of two) using a bitwise mask.
We also modify Lucet's stack overflow protection: Lucet emits conditional
checks to ensure that the stack does not overflow; these checks are rare and we
simply use \lfences.

We modify the springboard and trampoline transition functions in
the Lucet runtime.
Specifically, we add a single \lfence to each transition function since we must
disallow speculation from crossing the host-sandbox boundary.

The deterministic defenses for both \sysDesignOne and \sysDesignTwo---CBP
conversion and register interlocks---increase the cost of conditional
control flow.
To reduce the number of conditional branches, we thus enable explicit loop
unrolling flags when compiling the deterministic schemes.\footnote{
For simplicity, we do this in the Clang compiler when compiling applications to
Wasm and not in Lucet proper.
}
This is not necessary for the ASLR-based variants since they do not modify
conditional branches. 
Indeed, the ASLR variants are straightforward modifications to the dynamic
library loader used by the Lucet runtime:
Since all sandbox code is position independent, we just copy a new sandbox
instance's code and data pages to a new randomized location in memory.

We also made changes specific to each \sys scheme:

\paragraph{\sysDesignOne}
We augment the Cranelift code generation pass to
replace \mono{call} and \mono{return} instructions with the \sysDesignOne
separate stack instruction sequences and we mask pointers when they are unspilled from the stack.
For the deterministic variant of \sysDesignOne, we also
replace conditional branches with indirect jump instructions,
as described in \Cref{sec:sfi}.

To protect against \attackTwo attacks~(\S\ref{sec:SpectreOnWasm}), we flush the
BTB during the springboard transition into any sandbox.
Since this is a privileged operation, we implement this using a custom Linux
kernel module.

\paragraph{\sysDesignTwo}
In the \sysDesignTwo code generation pass, we place \mono{endbranch} instructions at
each indirect jump target in \wasm to enable \cet protection.
These indirect jump targets include switch table entries and functions which
may be called indirectly.
We also use this pass to emit the register interlocks for the deterministic variant
of \sysDesignTwo. 

We adapt the springboard and trampoline transition
functions to ensure that all uses of \mono{jmp}, \mono{call} and \mono{return}
conform to the requirements of \cet.
We furthermore use these transition functions to switch between the application
and sandbox \mpk domains.

Since \mpk blocks the application from accessing sandbox memory, we add
primitives that briefly turn off \mpk to copy memory into and out of sandboxes.
We implement these primitives using the \mono{rep} instruction prefix instead
of branching code, ensuring that the primitives are not vulnerable to Spectre attacks during
this window.

Finally, we add \cet support to both the Rust compiler---used to compile
Lucet---and to Lucet itself so that the resulting binaries are compatible with
the hardware.

\section{Evaluation}
\label{sec:eval}
\begin{figure*}[htb!]
  \begin{subfigure}{0.49\textwidth}
    \includegraphics[width=8.5cm]{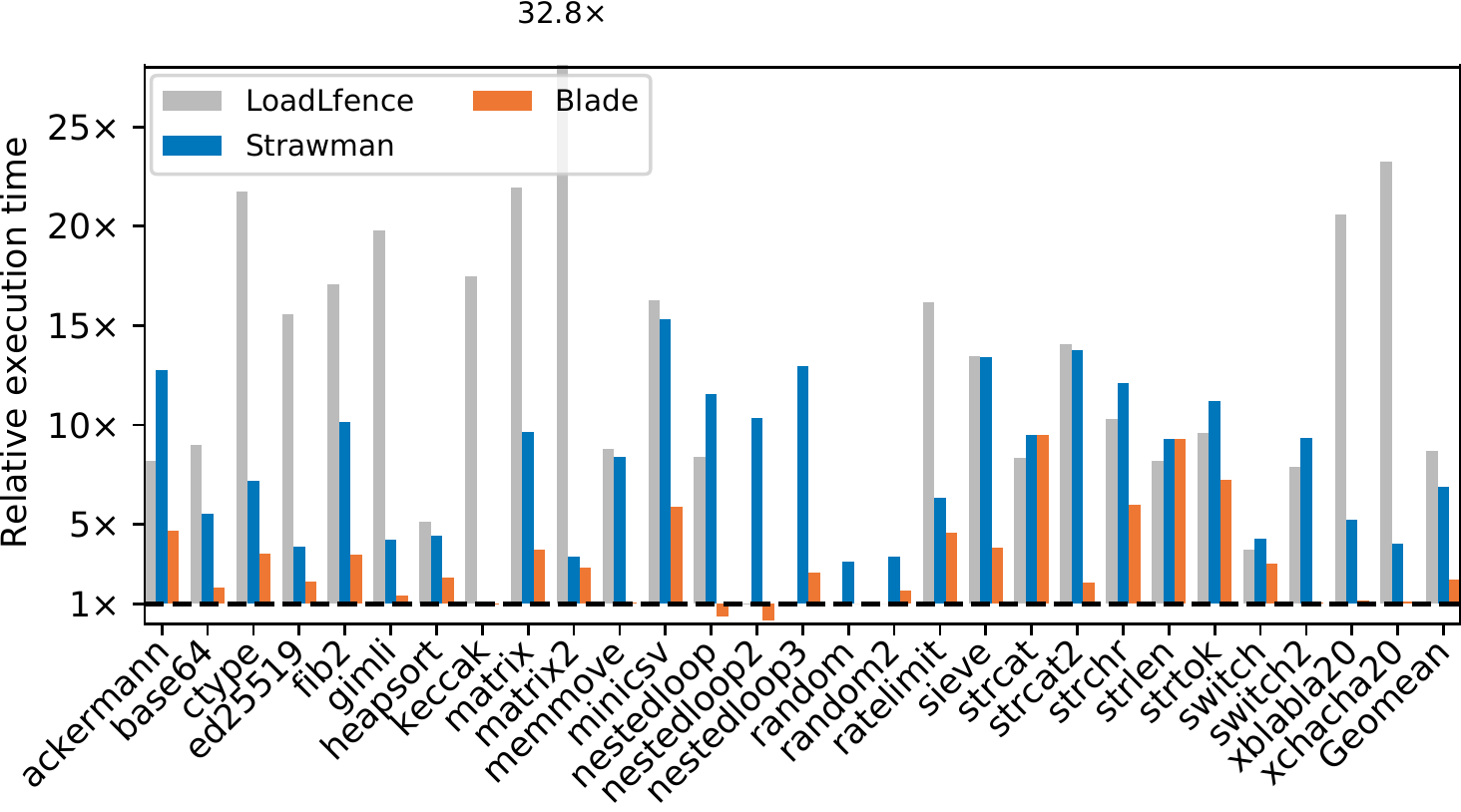}
    \caption{Fence scheme overhead on Sightglass}
    \label{fig:sightglass_fence}
  \end{subfigure}
  \begin{subfigure}{0.49\textwidth}
    \includegraphics[width=8.5cm]{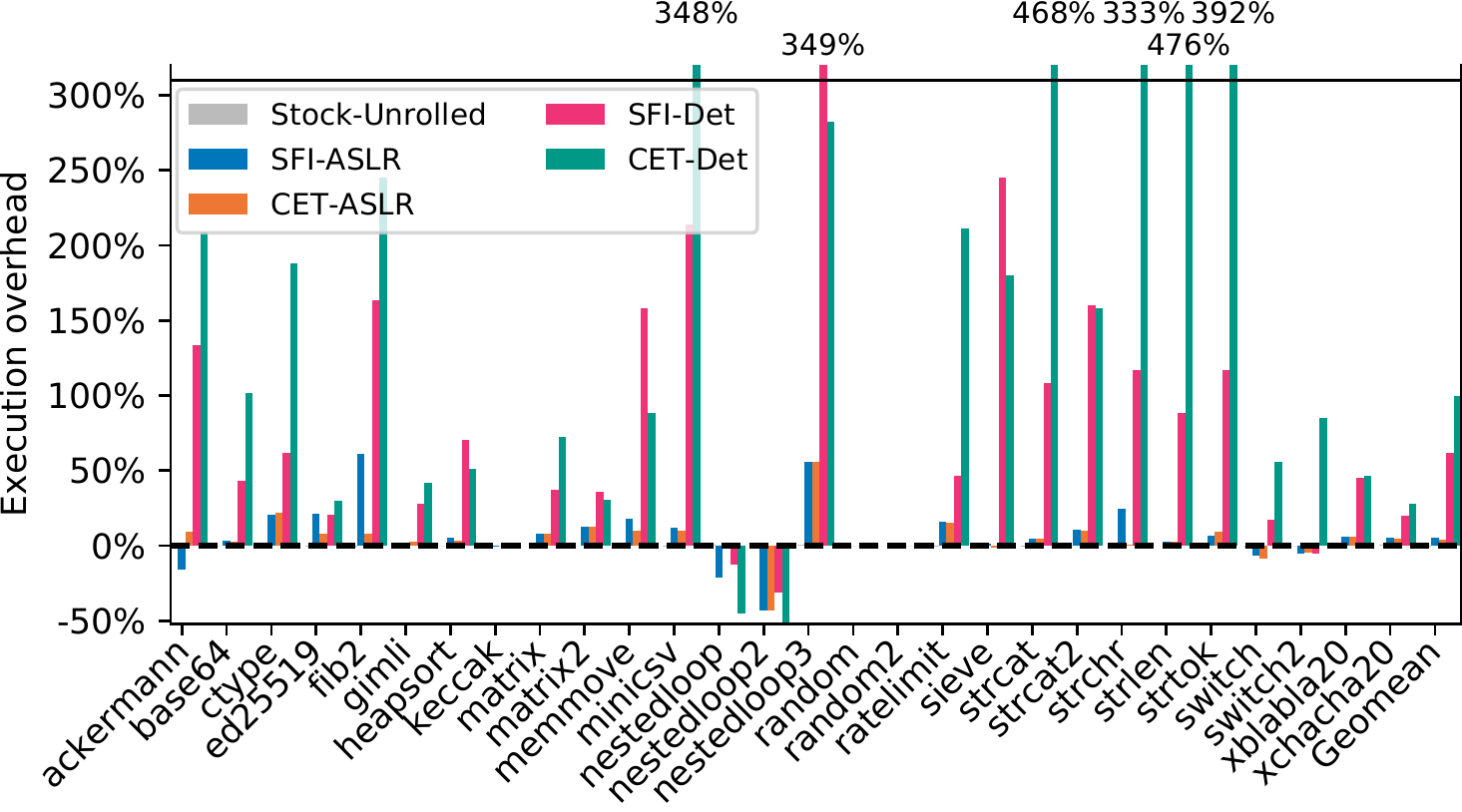}
    \caption{Swivel scheme overhead on Sightglass}
    \label{fig:sightglass}
  \end{subfigure}
  \caption{
    Performance overhead of \sys on the Sightglass benchmarks.
    (a) On Sightglass, the baseline schemes \loadfence, \sysStrawman, and
    \mincut  incur geomean overheads of \sgLoadfenceOverhead,
    \sgStrawmanOverhead, and \sgMincutOverhead respectively.
    (b) In contrast, the Swivel schemes perform much better where the ASLR
    versions of \sysDesignOne and \sysDesignTwo incur geomean
    overheads of \sgSfiASLROverhead and \sgCetASLROverhead respectively.
    With deterministic \attackTwo mitigations, these overheads are
    \sgSfiFullOverhead and \sgCetFullOverhead.
  }
  \label{fig:wasm_overhead_sightglass}
\end{figure*}
\begin{figure*}[htb!]
  \begin{subfigure}{0.49\textwidth}
    \includegraphics[width=8.5cm]{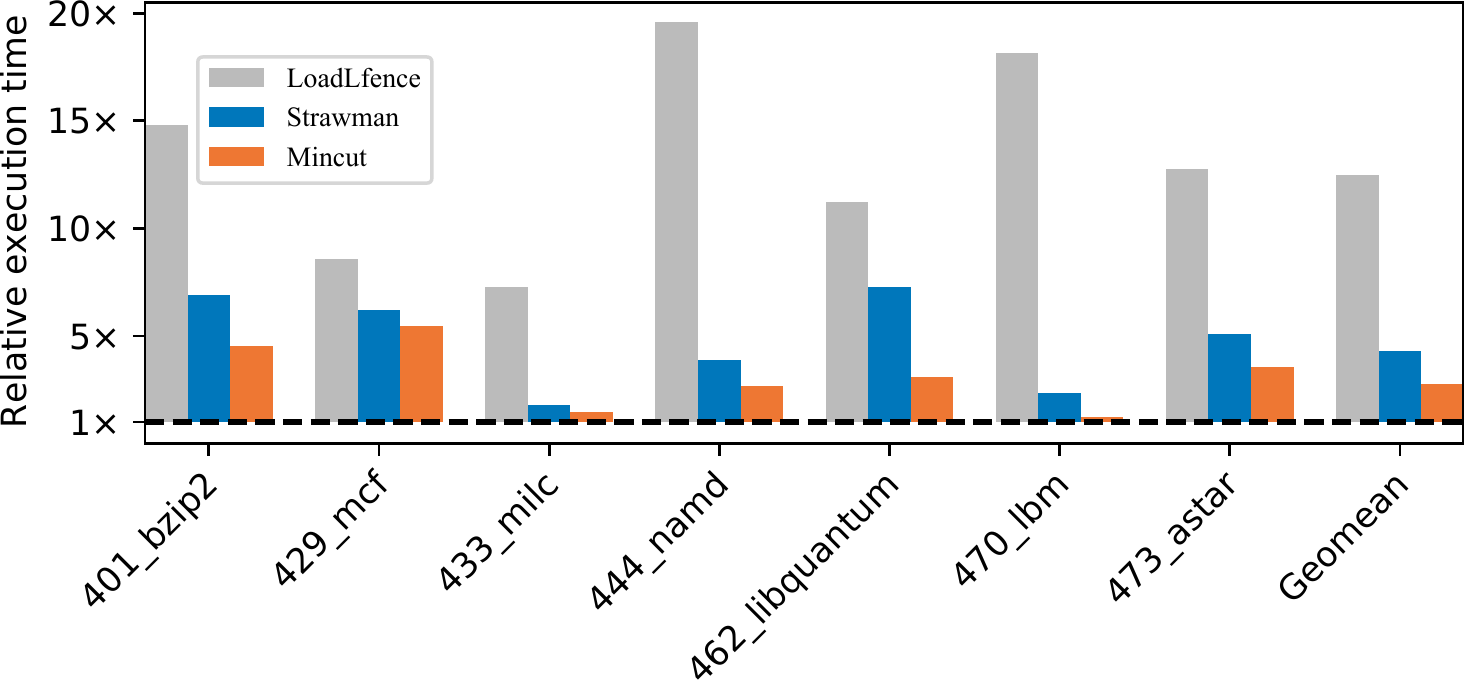}
    \caption{Fence scheme overhead on SPEC~2006}
    \label{fig:spec_fence}
  \end{subfigure}
  \begin{subfigure}{0.49\textwidth}
    \includegraphics[width=8.5cm]{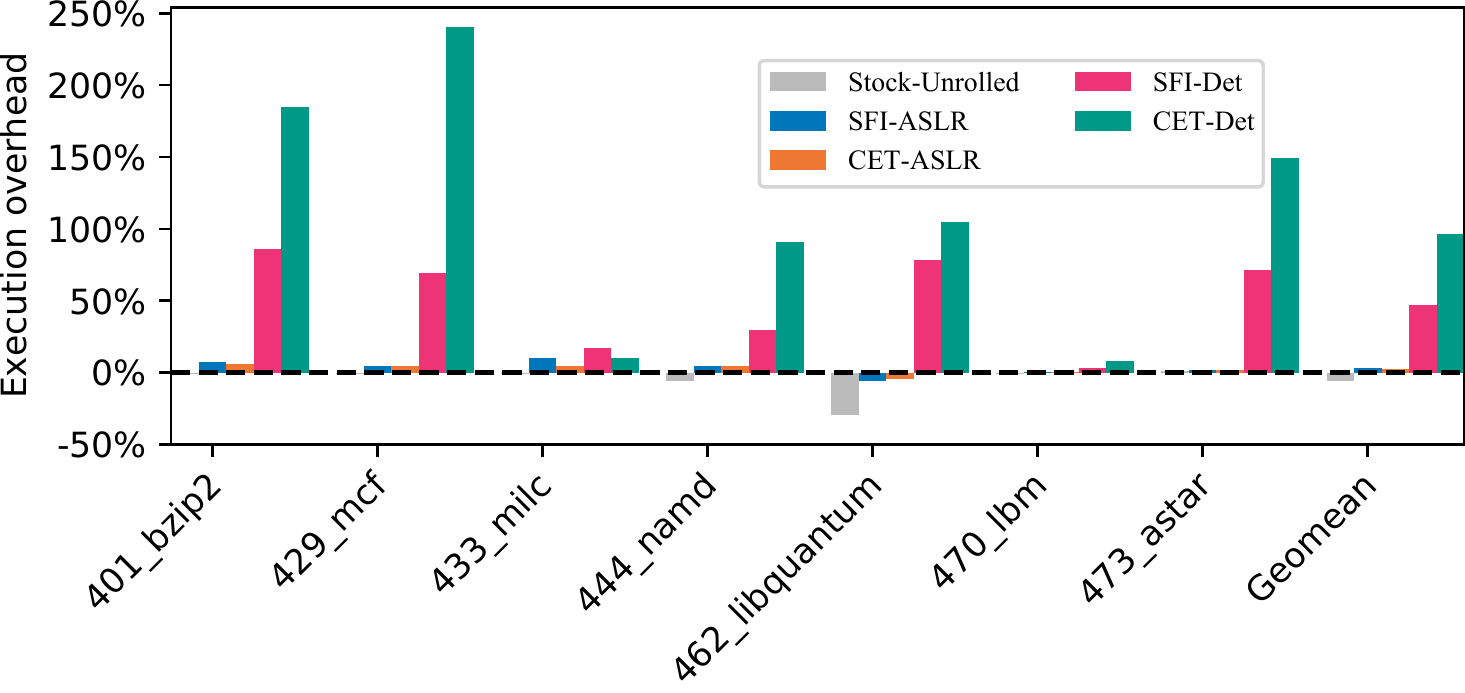}
    \caption{Swivel scheme overhead on SPEC~2006}
    \label{fig:spec}
  \end{subfigure}
  \caption{
    Performance overhead of \sys on SPEC~2006 benchmarks.
    (a) On SPEC~2006, the baseline schemes \loadfence, \sysStrawman, and \mincut
    incur overheads of \specLoadfenceMinOverhead--\specLoadfenceMaxOverhead,
    \specStrawmanMinOverhead--\specStrawmanMaxOverhead, and
    \specMincutMinOverhead--\specMincutMaxOverhead respectively.
    (b) In contrast, the Swivel schemes perform much better where the ASLR
    versions of \sysDesignOne and \sysDesignTwo incur overheads of at most
    \specSfiASLRMaxOverhead and \specCetASLRMaxOverhead respectively.
    With deterministic \attackTwo mitigations, these overheads are
    \specSfiFullMinOverhead--\specSfiFullMaxOverhead and
    \specCetFullMinOverhead--\specCetFullMaxOverhead respectively.
  }
  \label{fig:wasm_overhead_spec}
\end{figure*}

We evaluate \sys by asking four questions:
\begin{CompactItemize}
\item \textbf{What is the overhead of \wasm execution? (\S\ref{subsec:wasm_overhead})}
\sys's hardening schemes make changes to the code generated by the Lucet \wasm
compiler.
We examine the performance impact of these changes on Lucet's Sightglass
    benchmark suite~\cite{sightglass} and Wasm-compatible
    SPEC~2006~\cite{spec2006} benchmarks.

\item \textbf{What is the overhead of transitions? (\S\ref{subsec:transition_overhead})}
Swivel modifies the trampolines and springboards used to transition into and
out of Wasm sandboxes.
The changes vary across our different schemes---from adding {\lfence}s, or
flushing the BTB during one or both transition directions, to switching \mpk
domains.
We measure the impact of these changes on transition costs using a
microbenchmark.

\item \textbf{What is the end-to-end overhead of \sys? (\S\ref{subsec:application_overhead})}
We examine the impact of \wasm execution overhead and transition
overhead on a webserver that runs \wasm services.
We measure the impact of \sys protections on five different \wasm workloads
running on this webserver.

\item \textbf{Does \sys eliminate Spectre attacks? (\S\ref{sec:pocs})}
We evalute the security of \sys, i.e., whether \sys prevents sandbox breakout
and poisoning attacks, by implementing several
    proof-of-concept Spectre attacks.
\end{CompactItemize}

\paragraph{Machine setup}
We run our benchmarks on a 4-core, 8-thread Tigerlake CPU software
development platform (2.7GHz with a turbo boost of 4.2GHz) supporting the \cet
extension.
The machine has 16~GB of RAM and runs 64-bit Fedora~32 with the 5.7.0 Linux
kernel modified to include \cet support~\cite{cet-kernel}.
Our \sys modifications are applied to Lucet version \mono{0.7.0-dev}, which
includes Cranelift version \mono{0.62.0}.
We perform benchmarks on standard SPEC~CPU~2006, and Sightglass version
\mono{0.1.0}.
Our webserver macrobenchmark relies on the Rocket webserver version
\mono{0.4.4}, and we use \mono{wrk} version \mono{4.1.0} for testing.

\subsection{\wasm execution overhead}
\label{subsec:wasm_overhead}
We measure the impact of \sys's Spectre mitigations on \wasm performance in
Lucet using two benchmark suites:

\begin{CompactItemize}
  \item
The Sightglass benchmark suite~\cite{sightglass}, used by the Lucet compiler,
includes small functions such as cryptographic primitives (\mono{ed25519},
\mono{xchacha20}), mathematical functions (\mono{ackermann}, \mono{sieve}), and
common programming utilities (\mono{heapsort}, \mono{strcat}, \mono{strtok}).

\item
SPEC~CPU~2006 is a popular performance benchmark that includes various integer
and floating point workloads from typical programs.
We evaluate on only the subset of the benchmarks from SPEC~2006 that are
compatible with \wasm and Lucet.  This excludes programs written in Fortran,
programs that rely on dynamic code rewriting, programs that require more
than 4GB of memory which \wasm does not support, and programs that use
exceptions, \mono{longjmp}, or multithreading.\footnote{Some Web-focused \wasm
platforms support some of these features.
Indeed, previous academic work evaluates these benchmarks on
\wasm~\cite{notsofast}, but non-Web \wasm platforms including Lucet do not
support them.}
We note that \sys does \emph{not} introduce new incompatibilities with SPEC~2006
benchmarks; all of \sys's schemes are compatible with the same benchmarks as
stock Lucet.
\end{CompactItemize}

\paragraph{Setup}
We compile both Sightglass and the SPEC~2006 benchmarks with our modified Lucet
\wasm compiler and run them with the default settings.
Sightglass repeats each test at least 10 times or for a total of 100ms
(whichever occurs first) and reports the median runtime.
We compare \sys's performance overhead
with respect to the performance of the same benchmarks when using the stock
Lucet compiler.
For increased measurement consistency on short-running benchmarks, while
measuring Sightglass we pin execution to a single core and disable CPU
frequency scaling.\footnote{Tests were performed on June 18, 2020; see testing disclaimer~\ref{sec:app:testing-disclaimer}.}

\paragraph{Baseline schemes}
In addition to our comparison against Stock Lucet, we also implement three
known Spectre mitigations using {\lfence}s and compare against these as a
reference point.
First, we implement \loadfence, which places an \lfence after every load,
similar to Microsoft's Visual Studio compiler's ``Qspectre-load''
mitigation~\cite{mscvcspectre}.
Next, we implement \sysStrawman, a scheme which restricts code to sequential
execution by placing an \lfence at all control flow targets (at the start of
all linear blocks)---this is similar to the Intel compiler's ``all-fix-lfence''
mitigation~\cite{icc-spectre}.
Finally, we implement \mincut, an \lfence insertion algorithm suggested by
Vassena \etal\cite{blade} which uses a min-cut algorithm to minimize the
number of required {\lfence}s.
We further augment \mincut's \lfence insertion with several of our own
optimizations, including (1) only inserting a single \lfence per linear block;
and (3) unrolling loops to minimize branches, as we do for the register
interlock scheme and CBP conversions~(\S\ref{subsubsec:cbp-to-btb}).
Finally, to ensure that unrolling loops does not provide an unfair advantage,
we also present results for \mono{Stock-Unrolled}, which is stock Lucet with
the same loop unrolling as used in \sys's schemes.

\paragraph{Results}
We present the \wasm execution overhead of the various protection options
on the Sightglass benchmarks in \Cref{fig:sightglass}, and on the SPEC~2006
benchmarks in \Cref{fig:spec}.
The overheads of the ASLR versions of \sysDesignOne and \sysDesignTwo are
small: \sgSfiASLROverhead and \sgCetASLROverhead geomean overheads
respectively on Sightglass, and at most \specSfiASLRMaxOverhead (geomean:
\specSfiASLROverhead) and at most \specCetASLRMaxOverhead (geomean:
\specCetASLROverhead) respectively on SPEC.
The deterministic versions of \sys introduce modest
overheads: \sgSfiFullOverhead and \sgCetFullOverhead on Sightglass, and
\specSfiFullMinOverhead--\specSfiFullMaxOverhead (geomean:
\specSfiFullOverhead) and \specCetFullMinOverhead--\specCetFullMaxOverhead
(geomean: \specCetFullOverhead) on SPEC.
All four configurations outperform the baseline schemes by orders of magnitude:
\sysStrawman incurs geomean overheads of \sgStrawmanOverhead and
\specStrawmanOverhead on Sightglass and SPEC respectively, \loadfence incurs
\sgLoadfenceOverhead and \specLoadfenceOverhead overhead respectively, while
\mincut incurs \sgMincutOverhead and \specMincutOverhead respectively.

\paragraph{Breakdown}
Addressing CBP poisoning (CBP-to-BTB conversion in \sysDesignOne and register
interlocks in \sysDesignTwo) dominates the performance overhead of our
deterministic implementations.
We confirm this hypothesis with a microbenchmark: We measure the overheads of
these techniques individually on stock Lucet (with our loop
unrolling flags).
We find that the average (geomean) overhead of CBP conversion is
\sgPHTtoBTBOverhead on Sightglass and \specPHTtoBTBOverhead on SPEC.
The corresponding overheads for register interlocks are \sgInterlockOverhead
and \specInterlockOverhead.

Increasing loop unrolling thresholds does not significantly improve the
performance of stock Lucet (e.g., the speedup of our loop unrolling on stock
Lucet is \sgStockUnrollOverhead and \specStockUnrollOverhead on Sightglass and
SPEC, respectively).
It does impact the performance of our deterministic \sys variants though (e.g.,
we find that it contributes to a 15\%-20\% speed up).
This is not surprising since loop unrolling results in fewer conditional
branches (and thus reduces the effect of CBP conversions and register
interlocking).

To understand the outliers in \Cref{fig:wasm_overhead_sightglass} and
\Cref{fig:wasm_overhead_spec}, we inspect the source of the benchmarks.
Some of the largest overheads in \Cref{fig:wasm_overhead_sightglass} are on
Sightglass' string manipulation benchmarks, \eg \mono{strcat} and \mono{strlen}.
These microbenchmarks have lots of \emph{data-dependent loops}---tight loops
with data-dependent conditions---that cannot be unrolled at compile-time.
Since our register interlocking inserts a data dependence between the pinned
heap base register and the loop condition, this prevents the CPU from
speculatively executing instructions from subsequent iterations.
We believe that similar data-dependent loops are largely the cause for the
slowdowns on SPEC benchmarks, including \mono{429.mcf} and \mono{401.bzip2}.
Some of the other large overheads in Sightglass (e.g., \mono{fib2} and
\mono{nestedloop3}) are largely artifacts of the benchmarking suite:
These microbenchmarks test simple constructs like loops---and CBP-to-BTB
conversion naturally makes (almost empty) loops slow.

\subsection{Sandbox transition overhead}
\label{subsec:transition_overhead}
We evaluate the overhead of context switching.
As described in \Cref{sec:design}, \sys adds an {\lfence} instruction to
host-sandbox transitions to mitigate \attackOne attacks.
In addition to this:
\sysDesignOne flushes the BTB during each transition;
\sysDesignTwo, in deterministic mode, switches \mpk domains during each
transition;
and \sysDesignTwo, in ASLR mode, flushes the BTB in one direction and switches
\mpk domain in each transition.

We measure the time required for the host application to invoke a simple
no-op function call in the sandbox, as well as the time required for the
sandboxed code to invoke a permitted function in the application (\ie perform
a callback).
We compare the time required for \wasm code compiled by stock Lucet with the
time required for code compiled with our various protection schemes.
We measure the average performance overhead across 1000 such function call
invocations.\footnote{Tests were performed on June 18, 2020; see testing disclaimer~\ref{sec:app:testing-disclaimer}.}
These measurements are presented in
\Cref{tab:transitions}.

\begin{table}[t]
  \centering
  \footnotesize
  \begin{tabular} {p{5cm}|c|c}
    \toprule
    \textbf{\pbox{\textwidth}{Transition Type}} &
    \textbf{\pbox{\textwidth}{Function\\Invoke}} &
    \textbf{\pbox{\textwidth}{Callback\\Invoke}} \\
    \midrule
    Stock &
    \stockTransition &
    \stockTransitionCb \\

    \sysDesignOne (\lfence + BTB flush both ways) &
    \sfiTransition &
    \sfiTransitionCb \\

    \sysDesignTwo ASLR (\lfence + BTB flush one way + MPK) &
    \cetTransitionAslr &
    \cetTransitionAslrCb \\

    \sysDesignTwo deterministic (\lfence + MPK) &
    \cetTransition &
    \cetTransitionCb \\

    \bottomrule
  \end{tabular}
  \caption{
    Time taken for transitions between the application and sandbox---for
    function calls into the sandbox and callback invocations from the
    sandbox.
    \sys overheads are generally modest, with the deterministic variant of
    \sysDesignTwo in particular imposing very low overheads.
  }
  \label{tab:transitions}
\end{table}

\begin{table*}[t!]
  \centering
\begin{adjustbox}{max width=\textwidth}
\begin{tabular}{l|rrrr|rrrr|rrrr|rrrr}

\toprule
\multirow{2}{1cm}{\textbf{\pbox{\textwidth}{\sys Protection}}}
 & \multicolumn{4}{c|}{\bf\cdnTemplatedHTML}
 & \multicolumn{4}{c|}{\bf\cdnXMLtoJSON}
 & \multicolumn{4}{c|}{\bf\cdnJpgQuality}
 & \multicolumn{4}{c}{\bf\cdnHash}
\\\cline{2-17}

 &  \multicolumn{1}{c}{\textbf{ALat}} & \multicolumn{1}{c}{\textbf{TLat}} & \multicolumn{1}{c}{\textbf{Tput}} & \multicolumn{1}{c|}{\textbf{Size}} 
 &  \multicolumn{1}{c}{\textbf{ALat}} & \multicolumn{1}{c}{\textbf{TLat}} & \multicolumn{1}{c}{\textbf{Tput}} & \multicolumn{1}{c|}{\textbf{Size}}
 &  \multicolumn{1}{c}{\textbf{ALat}} & \multicolumn{1}{c}{\textbf{TLat}} & \multicolumn{1}{c}{\textbf{Tput}} & \multicolumn{1}{c|}{\textbf{Size}}
 &  \multicolumn{1}{c}{\textbf{ALat}} & \multicolumn{1}{c}{\textbf{TLat}} & \multicolumn{1}{c}{\textbf{Tput}} & \multicolumn{1}{c}{\textbf{Size}}
\\\midrule

  Stock (unsafe)
 & 20.8ms       & 42.1ms       & 4.81k        & 3.3MB
 & 186ms        & 228ms        & 531          & 3.2MB
 & 2.23s        & 2.93s        & 38.2         & 2.0MB
 & 424ms        & 532ms        & 230          & 3.6MB
\\\hline

\sysDesignOne ASLR
 & 124ms        & 137ms        & 803          & 3.9MB
 & 213ms        & 281ms        & 459          & 3.8MB
 & 2.31s        & 2.91s        & 36.9         & 2.2MB
 & 449ms        & 608ms        & 215          & 4.2MB
\\\hline

\sysDesignOne Det
 & 34.6ms       & 80.4ms       & 2.90k        & 4.2MB
 & 279ms        & 322ms        & 350          & 4.1MB
 & 3.01s        & 4.13s        & 26.4         & 2.9MB
 & 463ms        & 575ms        & 210          & 4.6MB
\\\hline

\sysDesignTwo ASLR
 & 111ms        & 123ms        & 898          & 3.4MB
 & 197ms        & 252ms        & 498          & 3.3MB
 & 2.30s        & 2.88s        & 37.0         & 2.0MB
 & 409ms        & 562ms        & 234          & 3.7MB
\\\hline

\sysDesignTwo Det
 & 28.7ms       & 66.3ms       & 3.50k        & 4.1MB
 & 291ms        & 328ms        & 338          & 4.0MB
 & 2.92s        & 3.81s        & 27.5         & 2.9MB
 & 459ms        & 570ms        & 211          & 4.4MB
\\\bottomrule

\end{tabular}
\end{adjustbox}

  \caption{
    Average latency (\textbf{ALat}), 99\% tail latency (\textbf{TLat}), average throughput 
    (\textbf{Tput}) in requests/second and binary files size (\textbf{Size}) for the
    webserver with different \wasm workloads
    (1k = $10^3$, 1m = $10^6$).
  }
  \label{tab:webserver_latthrough_1}
\end{table*}

\begin{table}[h]
  \centering
  \footnotesize
\begin{tabular}{l|rrrr}

\toprule
\multirow{2}{1cm}{\textbf{\pbox{\textwidth}{\sys Protection}}}
  & \multicolumn{4}{c}{\textbf{\cdnML}}
\\\cline{2-5}

  & \multicolumn{1}{c}{\textbf{ALat}} & \multicolumn{1}{c}{\textbf{TLat}} & \multicolumn{1}{c}{\textbf{Tput}} & \multicolumn{1}{c}{\textbf{Size}}
\\\midrule

Stock (unsafe)
 & 9.67s        & 13.1s        & 2.05         & 34.2MB
\\\hline

\sysDesignOne ASLR
 & 9.78s        & 13.9s        & 2.03         & 34.3MB
\\\hline

\sysDesignOne Det
 & 17.7s        & 28.3s        & 1.11         & 34.7MB
\\\hline

\sysDesignTwo ASLR
 & 9.82s        & 12.8s        & 2.02         & 34.2MB
\\\hline

\sysDesignTwo Det
 & 15.7s        & 24.9s        & 1.26         & 34.7MB
\\\bottomrule

\end{tabular}

  \caption{
    Average latency (\textbf{ALat}), 99\% tail latency (\textbf{TLat}), average throughput 
    (\textbf{Tput}) in requests/second and binary files size (\textbf{Size}) for the
    webserver for a long-running, compute-heavy \wasm workload
    (1k = $10^3$, 1m = $10^6$).
  }
  \label{tab:webserver_latthrough_2}
\end{table}

First, we briefly note that function calls in stock Lucet take much longer
than callbacks.
This is because the Lucet runtime has not fully optimized the function call
transition, as these are relatively rare compared to callback transitions,
which occur during every syscall.

In general, \sys's overheads are modest, with the deterministic variant of
\sysDesignTwo in particular imposing very low overheads.
Flushing the BTB does increase transition costs, but the overall effect of
this increase depends on how frequently transitions occur between the
application and sandbox.
In addition, flushing the BTB affects not only transition performance but
also the performance of both the host application and sandboxed code.
Fully understanding these overheads requires that we evaluate the
overall performance impact on real world
applications, which we do next.

\subsection{Application overhead}
\label{subsec:application_overhead}
We now evaluate \sys's end-to-end performance impact on a
webserver which uses \wasm to host isolated web services.

\paragraph{Setup}
For this benchmark, we use the Rocket webserver~\cite{rocket}, which can
host web services written as \wasm modules.
Rocket operates very similarly to webservers used in previous academic papers
exploring \wasm modules~\cite{hall2019execution, faasm} as well as frameworks
used by CDNs such as Fastly.
We measure the webserver's performance while hosting five different web
services with varying CPU and IO profiles.
These services perform the following five tasks respectively:
(1) expanding an HTML template;
(2) converting XML input to JSON output;
(3) re-encoding a JPEG image to change image quality;
(4) computing the SHA-256 hash of a given input;
and (5) performing image classification using inference on a pretrained neural
network.
We measure the overall performance of the webserver by tracking the average
latency, 99\% tail-latency, and throughput for each of the five web services.
We also measure the size of the \wasm binaries produced.\footnote{Tests were performed on June 18, 2020; see testing disclaimer~\ref{sec:app:testing-disclaimer}.}

\paragraph{Results}
Tables~\ref{tab:webserver_latthrough_1} and~\ref{tab:webserver_latthrough_2}
show results of the webserver measurements.
From the table, we see any of sys's schemes only reduce geomean throughput
(across all workloads) between \cdnMinOverhead and \cdnMaxOverhead.
\sys also modestly increases \wasm binary sizes, particularly with its
deterministic schemes, due to additional instructions added for
separate stack, CBP-to-BTB, and interlock mechanisms.

For long-running, compute-heavy \wasm workloads such as JPEG re-encoding and
image classification, \sys's performance overhead is dominated by \wasm
execution overhead measured in \Cref{subsec:wasm_overhead}.
Thus, on these workloads the ASLR versions of \sys perform much better than
the deterministic versions, as their \wasm execution overhead is lower.
On the other hand, for short-running workloads such as templated HTML, we
observe that the deterministic schemes outperform the ASLR schemes.
This is because \sys's ASLR implementation must remap and \mono{memcpy} the
sandbox code pages during sandbox creation, effectively adding a fixed
overhead to each request.
For short-running requests, this fixed per-request cost dominates overall
overhead.
In contrast, Stock Lucet and \sys's deterministic schemes take advantage of
shared code pages in memory to create sandboxes more rapidly, incurring lower
overhead on short-running requests.

\subsection{Security evaluation}
\label{sec:pocs}
To evaluate the security of \sys, we implement several Spectre attacks in
\wasm and compile this attack code with both stock Lucet and \sys.
We find that stock Lucet produces code that is vulnerable to Spectre, i.e., our
proof of concept attacks (POCs) can be used to carry out both breakout and
poisoning attacks, and that \sys mitigates these attacks.

\paragraph{Attack assumptions}
Our attacks extend Google's Safeside~\cite{safeside} suite and, like the
Safeside POCs, rely on three low-level instructions: The \mono{rdtsc}
instruction to measure execution time, the \mono{clflush} instruction to evict
a particular cache line, and the \mono{mfence} instruction to wait for pending
memory operations to complete.
While these instructions are not exposed to \wasm code by default, we expose
these instructions to simplify our POCs.
Additionally, for cross \wasm module attacks, we manually specify the locations
where Wasm modules are loaded to simplify the task of finding partial address
collisions in the branch predictor.

Our simplifications are not fundamental and can be removed in an end-to-end
attack.
Previous work, for example, showed how to construct precise
timers~\cite{schwarz2017fantastic,frigo2018grand}, and how to control cache
contents~\cite{vila2019theory} in environments like JavaScript where these
instructions are not directly exposed.
The effects of the \mono{mfence} instruction can be achieved by
executing \mono{nop} instructions until all memory operations are drained.
And, in the style of heap and JIT spraying attacks~\cite{sintsov2010jit}, we
can increase the likelihood of partial address collision by deploying hundreds
to thousands of modules on the FaaS platform.

\paragraph{POC~1: Sandbox breakout via in-place Spectre-PHT}
Our first POC adopts the original Spectre-PHT bounds-check bypass
attack~\cite{Kocher2019spectre} to \wasm.
As mentioned in Section~\ref{sec:SpectreOnWasm}, in \wasm, indirect function
calls are expressed as indices into a function table.
Hence, the code emitted for the \asm{call_indirect} instruction performs a
bounds check, to ensure that the function index is within the bounds of the
table, before performing the lookup and call.
By inducing a misprediction on this check, our POC can read beyond the function
table boundary and treat the read value as a function pointer.
This effectively allows us to jump to any code location and speculatively
bypass \wasm's CFI (and thus isolation).
We demonstrate this by jumping to a host function that returns bytes of a
secret array.

\paragraph{POC~2: Sandbox breakout and poisoning via out-of-place Spectre-BTB}
Our second POC adopts the out-of-place Spectre-BTB attack of
Canella et al.~\cite{Canella2019} to Wasm.
Specifically, we mistrain an indirect jump in a victim or attacker-controlled module
by training a congruent indirect jump instruction in another attacker-controlled module.
We train the jump to land on a gadget of our choice.
To demonstrate the feasibility of a sandbox poisoning attack, we target a
double-fetch leak gadget.
To demonstrate a sandbox breakout attack, we jump in the middle of a basic
block to a memory load, skipping \wasm's heap bounds checks.

\paragraph{POC~3: Poisoning via out-of-place Spectre-RSB}
Our third POC compiles the Spectre-RSB attack from the Google Safeside
project~\cite{safeside} to Wasm.
This attack underflows the RSB to redirect speculative control flow.
We use this RSB underflow behavior to speculatively ``return'' to a gadget
that leaks module secrets.
We run this attack entirely within a single \wasm module.
However, on a FaaS platform this attack can be used across modules when the
FaaS runtime interleaves the execution of multiple modules, similar to the
Safeside cross-process Spectre-RSB attack.

\paragraph{Results}
We developed our POCs on a Skylake machine (Xeon Platinum 8160) and then tested
them on both this machine and the Tiger Lake \cet development platform we used
for our performance evaluation.
We found that stock Lucet on the Skylake machine was vulnerable to all three
POCs while \sysDesignOne, both the ASLR and deterministic versions, were not
vulnerable.
On the Tiger Lake machine, we found that stock Lucet was vulnerable to POC~3
while \sysDesignOne and \sysDesignTwo, both the ASLR and deterministic
versions, were not.
Although the Tiger Lake CPU is documented to be vulnerable to
all three Spectre variants~\cite{intelMitigationList}, we did not successfully
reproduce POC~1 and POC~2 on this machine.
Getting these attacks to work may require reverse engineering the branch
predictors used on these new CPUs. We thus leave the extensions of our
POCs to this microarchitecture to future work.

\section{Limitations and discussion}
\label{sec:discussion}

In this section, we cover some of the current limitations of \sys, briefly
mention alternate design points, and address the generality of our
solutions.

\subsection{Limitations of \sys}
We discuss some limitations of \sys, both in general and for our
implementation in particular.

\paragraph{Implementation limitations}
For this paper, we have simplified some of the implementation
details for \sysDesignTwo to reduce the engineering burden of modifying
multiple compiler toolchains and standard libraries while still providing
accurate performance evaluations.
First, we do not ensure that interlock labels are unique to each linear block,
but rather reuse interlock labels;
while unique labels are critical for security, previous works have extensively
demonstrated the feasibility of assigning unique labels~\cite{cfi-survey}.
Our goal was to measure the performance of the instruction sequences for
interlock assignment (64-bit conditional moves) and checking (64-bit conditional checks).

Next, when disabling \mpk protections in \sysDesignTwo~(\S\ref{sec:cet}) in the host
calls, we must avoid using indirect branches; while we follow this principle
for hostcalls we expose (e.g., when marshaling data for web server requests), we
do not modify existing standard library hostcalls.
These additional modifications, while straightforward, would require significant
engineering effort in modifying the standard library (libc) used by \wasm.
Finally, we did not implement the required guard pages in the lower 4GB of
memory in our prototype of deterministic \sysDesignTwo.
Prior work~\cite{isboxing} has shown how to reserve the bottom 4GB of
memory---and that this does not impact performance.

\paragraph{Secretless host}
\label{subsec:host-secrets}
\sys assumes that the host (or runtime) doesn't contain secret information.
This assumption is sensible for some applications: in the CDN use case, the CDN
part of the process is lightweight and exists only to coordinate with the
sandboxes.
But not all.
As a counter-example, the Firefox web browser currently uses
\wasm to sandbox third-party libraries written in C/C++~\cite{rlbox,rlbox-blog}.
We could use \sys to ensure that Firefox is secure from Spectre attacks
conducted by a compromised third-party libraries.
To protect secrets in the host (Firefox), we could either place
the secrets into a separate \wasm sandbox, or apply one of our proposed
CBP protections to the host (e.g., CBP-to-BTB or interlocks).

\paragraph{Hyperthreading}
The only scheme in \sys that supports hyperthreading is the deterministic
\sysDesignTwo.
Alternately, instead of disabling hyperthreading, Intel suggests relying on
single-threaded indirect branch predictors (STIBP) to prevent a co-resident
thread from influencing indirect branch predictions~\cite{IntelSpeculation}.
STIBP could allow any \sys scheme to be used securely with hyperthreading.

\subsection{Other leakages and transient attacks} \label{sec:otherattacks}
\sysDesignTwo allows victim code to run with poisoned predictors but prevents
exfiltration via the data cache.
This, unfortunately, means that attackers may still be able to leak victim data
through other microarchitectural channels (\eg port contention or the
instruction cache~\cite{Kocher2019spectre}).
\sysDesignOne does not have this limitation; we can borrow techniques from
\sysDesignOne to eliminate such leaks (e.g., flushing the BTB).

The CPU's memory subsystem may also introduce other transient execution attacks.
Spectre-STL~\cite{Horn2018spectre4} can leak stale data (which may belong
to another security domain) before a preceding store could overwrite this data
due to speculative dependency checking of the load and the preceding stores.
\sys does not address Spectre-STL. However, Spectre-STL can been mitigated
through speculative store bypass disable~(SSBD)~\cite{ssbd}, which imposes a
small performance overhead (less than 5\% on most
benchmarks~\cite{phoronix-ssbd}) and is already enabled by default on most
systems.

Meltdown~\cite{Lipp2018meltdown} could leak privileged kernel memory from
userspace.
Variants of the Meltdown attack, \eg microarchitectural data sampling (MDS), 
can be used to leak stale data from several microarchitectural
elements~\cite{Intel2019MDS,Canella2019Fallout,Schwarz2019ZL,moghimi2020Medusa,VanSchaik2019RIDL}.
Load Value Injection (LVI) exploits the same microarchitectural features as
Meltdown to inject data into the microarchitectural state~\cite{vanbulck2020lvi}.
More recent Intel CPUs (e.g., Tiger Lake) are designed to be resilient against
this class of attacks~\cite{intelMitigationList}, and we believe that these
attacks can efficiently be mitigated in hardware.
For this reason, recent research into secure speculation and architectural
defense against transient execution attacks are mostly focused on the Spectre class
of issues~\cite{guarnieri2018spectector,RFC,retpoline,wang2018oo7}.

For legacy systems, users should apply the latest microcode and software
patches to mitigate Meltdown and variants of MDS~\cite{kernelPTI,intelMitigationList}.
For variants of MDS that abuse hyperthreading on legacy systems,
Intel suggests safe scheduling of sibling CPU threads~\cite{Intel2019MDS}.
Since \wasm restricts what instructions are allowed in a \wasm module, this
makes some MDS attacks more challenging to execute.
For instance, \wasm modules cannot use \Intel TSX transactions or access
non-canonical or kernel addresses that are inherent to some of the MDS variants~\cite{TAA}.
LVI requires fine-grain control over inducing faults or microcode assists,
which is not available at the \wasm level.
Some legacy systems may still be vulnerable to LVI; however, the feasibility
of LVI attacks outside the Intel SGX environment is an open research
question~\cite{vanbulck2020lvi}.

\subsection{Alternate design points for \sys}
We next describe alternate designs for \sys and discuss the trade-offs of our
design choices.

\paragraph{CBP-to-BTB conversion in \sysDesignTwo}
Since register interlocking is more expensive than CBP-to-BTB conversion, a
reader may wonder whether the deterministic \sysDesignTwo could more
efficiently protect against \attackTwo using CBP-to-BTB conversion.
Unfortunately, since \sysDesignTwo does not flush the BTB in both directions,
CBP-to-BTB conversion is not sufficient to fully mitigate \attackTwo.
In addition to CBP-to-BTB conversion, \sysDesignTwo would also need to
use interlocking (without additional performance gain) or flush the BTB
both ways. In the latter case, we might as well use \sysDesignOne,
as the main advantage of using \cet in \sys is to avoid flushing the BTB.

\paragraph{Interlocking in \sysDesignOne}
Likewise, one may wonder about the benefits of using interlock in \sysDesignOne.
Unfortunately, for interlock to be useful, it requires hardware support.
First, we require \mpk to ensure that a sandbox can't confuse the host into
accessing (and then leaking) another sandbox's data.
Second, we require the \cet \mono{endbranch} instruction to
ensure that the sandbox cannot use BTB entries leftover from the host.

\paragraph{Partitioning shared resources}
A different approach to addressing Spectre would be to partition differ
hardware structures to ensure isolation.
For example, for the CBP, one approach would be to exploit the indexing
mechanism of branch predictors such that each sandbox uses an isolated
portion of the CBP.
Unfortunately doing this on existing CPUs is hard: superscalar CPUs use
complex predictors with multiple indexing functions,
and without knowledge of the underlying microarchitecture, we were unable to
experimentally find a way to partition the CBP.

Alternately, we could mitigate \attackThree---or even \attackTwo---attacks by
partitioning CPU cores, and preventing an attacker from running on the same core
as their victim.
This approach protects against \attackTwo and \attackThree, since branch
predictors are per-physical-core.
We tried this. Specifically, we implemented this mitigation in the \attackThree
context and measured its performance.
Unfortunately, requiring a core transition during every springboard and
trampoline is detrimental to performance, and this scheme was not competitive
with our chosen implementation.

\subsection{Generalizing \sys}
\sys's techniques are not specific to the Lucet compiler or runtime.
Our techniques can be applied to other \wasm compilers and runtimes, including
the just-in-time \wasm compilers used in the Chrome and Firefox browsers.

Our techniques can also be adopted to other software-based fault isolation
(SFI) compilers~\cite{tan-sfi-survey}.
Adopting \sys to the Native Client (NaCl) compiler~\cite{nacl, nacl64}, for
instance, only requires only a handful of changes.
For example, we wouldn't even need to add linear blocks: NaCl relies on
instruction bundles---32-byte aligned blocks of instructions---which are more
restrictive than our linear blocks (and satisfy our linear block invariants).

More generally, \sys can be adopted to other sandboxed languages and runtimes.
JavaScript just-in-time compilers are a particularly good fit.
Though JavaScript JITs are more complex than \wasm compilers, they
share a similar security model (e.g., JavaScript in the browser is untrusted)
and, in some cases, even share a common compilation pipeline.
For example, Cranelift---the backend used by Lucet and \sys---was designed to
replace Firefox's JavaScript and Wasm backend
implementations~\cite{cranelift-firefox}, and thus could transparently benefit
from our mitigations.
Beyond Cranelift, we think that adopting our linear blocks and code page
ASLR is relatively simple (e.g., compared to redesigning the browser to deal
with Spectre) and could make JavaScript Spectre attacks significantly more
difficult.

\subsection{Implementation bugs in \wasm}
Lehmann et al.~\cite{wasm-bugs} showed that some \wasm compilers and
runtimes, like prior SFI toolchains~\cite{tan-sfi-survey}, contain
implementation bugs.\footnote{
  They also show that C memory safety bugs are still present within the Wasm
  sandbox---this class of bugs is orthogonal and cannot alone be used to to
  bypass Wasm's isolation guarantees.
}
For example, they showed that some Wasm runtimes fail to properly separate
the stack and heap.
Though they did not identify such bugs in Lucet, these classes of bugs are
inevitable---and, while identifying such bugs is important, this class of bugs
is orthogonal and well-understood in the SFI literature (and addressed, for
example, by VeriWasm~\cite{veriwasm}).
We focus on addressing Spectre attacks, which can fundamentally undermine the
guarantees of even bug-free \wasm toolchains.

\subsection{Future work}

\sys's schemes can benefit from extensions to compiler toolchains as well as
hardware to both simplify its mitigations and improve performance.
We briefly discuss some possible extensions and their benefits below.

\subsubsection{Compiler toolchain extensions}

We describe two performance optimizations for the \sysDesignTwo deterministic
scheme, and a way to improve the security of \sys's ASLR schemes.

\paragraph{Data dependent loops}
As discussed in Section~\ref{subsec:wasm_overhead}, the \sysDesignTwo
deterministic scheme imposes the greatest overheads in programs with
data-dependent loops---e.g., programs that iterate over strings or
linked lists (which loop until they find a null element).
\sys effectively serializes iterations of such data-dependent loops.
We expect that many other Spectre mitigation (see Section~\ref{sec:related}),
like speculative taint tracking~\cite{yu2019speculative},
would similarly slow down such programs.

One way to speed up such code is to replace the data-dependent loops with a
code sequence that first counts the expected number of iterations~($N$),
executes an \lfence, and then runs the original loop body for $N$ iterations.
This would introduce only a single stall in the loop and eliminate the
serialization between loop iterations.

\paragraph{Compiler secret tracking}
\sys currently assumes all locations in memory contain potentially secret data.
However, several works (e.g.,~\cite{wang2018oo7}) have proposed tracking
secrets in compiler passes.
This information can be used to optimize the \sysDesignTwo deterministic
scheme. In particular, any public memory access can be hoisted above the
register interlock to allow the memory to be accessed (and ``leaked'')
speculatively.

\paragraph{Software diversity}
\sys's ASLR variants randomize code pages.
We could additionally use software diversity to increase the entropy of our
probabilistic schemes~\cite{software-diversity, ilr}.
Software diversity techniques (e.g., \mono{nop} insertion) are
cheap~\cite{nop-diversity}, and since they do not affect the behavior of
branches, they can be used to specifically mitigate out-of-place Spectre-BTB
and Spectre-PHT attacks.

\subsubsection{Hardware extensions}
Hardware extensions can make \sys faster and simpler.

\paragraph{CBP flushing}
\sysDesignOne schemes rely on ASLR or CBP-to-BTB conversion to protect the CBP.
However, hardware support for CBP flushing could significantly speed up \sys.
Alternatively, hardware support for tagging predictor state (e.g., host code
and sandbox code) would allow \sys to isolate the CBP without flushing.

\paragraph{Dedicated interlock instructions}
The register interlocking used in deterministic \sysDesignTwo requires
several machine instructions in each linear block in order to assign and
check labels.
Dedicated hardware support for these operations could reduce
code bloat.

\paragraph{Explicit BTB prediction range registers}
The \sysDesignTwo deterministic scheme allocates unique 64-bit labels to each
linear block, which do not overlap across sandbox instances.
We could simplify and speed up this scheme with a hardware extension that can
be used to limit BTB predictions to a range of addresses.
With such an extension, \sys could set the prediction range during each
transition into the sandbox (to the sandbox region) and ensure that the BTB
could only predict targets inside the sandbox code pages.
This would eliminate out-of-place BTB attacks---and, with linear blocks, it
would eliminate breakout attacks in \wasm.
Finally, this would reduce code size: it would allow us to to reduce block
labels to, for example, 16 bits (since we only need labels to be unique within
the sandbox).

\section{Related work}
\label{sec:related}

We give an overview of related work on mitigating Spectre
attacks by discussing microarchitectural proposals, software-based approaches
for eliminating Spectre gadgets, and previous approaches based on CFI or \mpk.

\paragraph{Thwarting covert channels}
Several works~\cite{kiriansky2018dawg,khasawneh2019safespec,barber2019specshield,yan2018invisispec,saileshwar2019cleanupspec}
propose making microarchitectural changes to block, isolate, or remove the
covert channels used to transfer transient secrets to architectural states.
For example, \emph{SafeSpec}~\cite{khasawneh2019safespec} proposes a
speculation-aware memory subsystem which ensures that microarchitectural
changes to the cache are not committed until predictions are validated.
Similarly, \emph{CleanupSpec}~\cite{saileshwar2019cleanupspec} proposes
an undo logic for the cache state. 
Although these approaches remove the attacker's data leakage channel, they do
not address the root cause of Spectre vulnerabilities.
In contrast, \sys works with no hardware changes.

\paragraph{Safe speculation}
Intel has introduced hardware support to mitigate Spectre-BTB across separate
address spaces~\cite{IntelSpeculation,intelMitigationList}.
Specifically, the Indirect Branch Predictor Barrier (IBPB) allows the BTB to
be cleared across context switches, while Single Thread Indirect Branch
Predictors (STIBP) ensure that one thread's BTB entries will not be affected
by the sibling hyperthread.
These mitigations can be used by the OS as a coarse-grained mechanism for
safe speculation, but only apply to Spectre-BTB and have not been widely
adopted due to performance overhead~\cite{koruyeh2019speccfi}.

Other works propose microarchitectural changes to allow the software to
control speculation for security-critical
operations~\cite{taram2019context,weisse2019nda} or certain memory
pages~\cite{schwarz2020context,li2019conditional}.
Separately, STT~\cite{yu2019speculative} proposes speculative taint tracking
within the microarchitecture.
However, unlike \sys, these approaches require significant hardware changes
and do not offer a way to safely run code on existing CPUs.

\paragraph{Eliminating Spectre gadgets}
Another way to mitigate Spectre attacks is by inserting a
barrier instruction (\eg \lfence), which blocks speculative
execution~\cite{mscvcspectre,IntelSpecMit,AMDSoftwareTechnique}.
However, as we evaluated in \Cref{subsec:wasm_overhead}, insertion of
\lfence has a performance impact on the entire CPU pipeline
and undercuts the performance benefit of out-of-order and speculative execution.
In contrast, \sys makes little to no use of \lfence.

An optimized approach is to replace control flow instructions with
alternate code sequences that are safe to execute speculatively.
For instance, speculative load hardening (SLH) replaces conditional bounds
checks with an arithmetized form to avoid Spectre-PHT~\cite{RFC}.
Indeed, \sys uses SLH to protect the bounds checks for indirect call tables and
switch tables (\S\ref{subsec:linear_blocks}).
Alternatively, Oleksenko et al.~\cite{oleksenko2018you} propose inserting
artificial data dependencies between secret operations and pipeline serialization
instructions.
Finally, the \emph{retpoline} technique~\cite{retpoline} replaces indirect
branches with a specific code sequence using the \mono{ret}
instruction to avoid Spectre-BTB.
To reduce the overhead of such code transformations, researchers have
proposed several techniques to automatically locate Spectre
gadgets~\cite{wang2018oo7,guarnieri2018spectector,cauligi2020foundations} and
apply mitigations to risky blocks of code.
However, these techniques have to handle potential false positives or negatives; in
contrast, \sys focuses on defending against all possible Spectre attacks
from untrusted code by applying compile-time mitigations.

\paragraph{Speculative CFI}
\emph{SpecCFI}~\cite{koruyeh2019speccfi} has proposed hardware support
for speculative and fine-grained control-flow integrity (CFI), which
can be used to protect against attacks on indirect branches.
In comparison, \sysDesignTwo uses \cet, which only supports coarse-grained
CFI with speculative guarantees~\cite{shanbhogue2019security}.
\emph{Venkman}~\cite{venkman} uses a technique similar to \sys's
linear blocks to ensure that indirect branches always reach a
barrier instruction (\eg \lfence) by applying alignment to bundles similar to
classical software fault isolation~\cite{wahbe-sfi}.
In contrast, \sys is a fence-free approach that preserves the performance
benefits of speculative execution.

\paragraph{Intra-process isolation using \mpk}
Jenkins \etal~\cite{jenkins2020ghostbusting} propose to provide intra-process
Spectre protection using \mpk.
They use \mpk to create separate isolation domains and use the relationship
between the code and secret data to limit speculative accesses.
However, since \mpk only provides 16 domains, relying fully on \mpk to isolate
many sandbox instances is infeasible for the CDN \wasm use case we consider.

\section{Conclusion}
\label{sec:conclusion}
This work proposes a framework, \sys, which provides strong in-memory
isolation for \wasm modules by protecting against Spectre attacks.
We describe two \sys designs: \sysDesignOne, a software-only approach which
provides mitigations compatible with existing CPUs, and \sysDesignTwo, which
leverages \cet and \mpk.
Our evaluation shows that versions of \sys using ASLR incur low performance
overhead (at most \specASLRMaxOverhead on compatible SPEC~2006 benchmarks),
demonstrating that \sys can provide strong security guarantees for \wasm
modules while maintaining the performance benefits of in-process
sandboxing.

\ifAnon
\else
\section*{Acknowledgment}
We thank Johnnie Birch, Jonathan Foote, Dan Gohman, Pat Hickey, Tyler McMullen,
Jan de Mooij, Vedvyas Shanbhogue, Jared Stark, Luke Wagner, and Andy Wortman for insightful discussions.
We thank Devdatta Akhawe and the anonymous reviewers for their valuable
comments for improving the quality of this paper.
We would also like to thank Hongjiu Lu and Yu-cheng Yu for their support on the \cet infrastructure.
This work was supported in part by gifts from Cisco, Fastly, Mozilla, and by
the CONIX Research Center, one of six centers in JUMP, a Semiconductor Research
Corporation (SRC) program sponsored by DARPA, by the NSF under grant
numbers CCF-1918573, CNS-1814406, CAREER CNS-2048262, and by NSF/Intel under
grant number CCF-1823444.
\fi

\ifUsenix
{
  \fontsize{7}{8}\selectfont
  \setlength{\bibsep}{2pt}
  \bibliographystyle{abbrv}
  \bibliography{local}
}
\else
  \bibliographystyle{IEEEtranS}
  {\footnotesize
  \bibliography{IEEEabrv,local}
  }
\fi

\appendix
\section{Appendix}

\subsection{Brief introduction to CET and MPK}
\label{app:cet}

\paragraph{CET} \cet is an instruction set architecture extension that helps 
prevent Return-Oriented Programming and Call/Jmp-Oriented Programming
via use of a \emph{shadow stack}, and \emph{indirect branch tracking} 
(IBT).
The shadow stack is a hardware-maintained stack used
exclusively to check the integrity of return addresses on the program stack.
To ensure the shadow stack cannot be tampered with, it is inaccessible via 
standard load and store instructions.
The IBT allows the enforcement of coarse-grained control flow integrity 
(CFI)~\cite{cfi} via a branch termination instruction, \mono{endbranch}.
Binaries that wish to use IBT place the \mono{endbranch} at all valid indirect 
jump targets.
If an indirect jump instruction lands on any other instruction, the CPU
reports a control-flow protection fault.
Additionally, the IBT also supports a \emph{legacy bitmap}, which allows 
programs to demarcate which code pages have IBT checking enabled.

Importantly, \cet guarantees that any shadow stack mismatches observed during 
speculative execution of \mono{return} instruction immediately halts further 
speculative execution.
Similarly, any indirect jump during speculative execution from an IBT enabled 
code page to a page with IBT disabled also halts speculation.

\paragraph{MPK} \mpk uses four bits in each page-table entry to assign one of sixteen 
"keys" to any given memory page, allowing for 16 different memory domains.
User mode instructions \mono{wrpkru} and \mono{rdpkru} allow setting read and 
write permissions for each of these domains on a per-thread basis.
\mpk thus allows a process to partition its memory and selectively 
enable/disable read and write access to any of regions without invoking the 
kernel functions or switching page tables.

Importantly, \mono{wrpkru} does not execute speculatively - memory accesses
affected by the PKRU register will not execute (even speculatively) until all
prior executions of \mono{wrpkru} have completed execution and updated the PKRU 
register and are also resistant to Meltdown style attacks~\cite{IntelSpecMit}.

\subsection{Testing Disclaimer}
\label{sec:app:testing-disclaimer}
Since we use a software development platform provided by Intel, we include the
following disclaimer from Intel:

\begin{myleftbar}
  \small
  Software and workloads used in performance tests may
have been optimized for performance only on Intel microprocessors. Performance
tests, such as SYSmark and MobileMark (in this paper SPEC CPU
2006 and Sightglass), are measured using specific computer systems, components,
software, operations and functions. Any change to any of those factors may cause
the results to vary. You should consult other information and performance tests
to assist you in fully evaluating your contemplated purchases, including the
performance of that product when combined with other products. For more complete
information visit www.intel.com/benchmarks. Performance results are based on
testing as of dates shown in configurations and may not reflect all publicly
available updates. See backup for configuration details. No product or component
can be absolutely secure. Your costs and results may vary. Intel technologies
may require enabled hardware, software or service activation.
\end{myleftbar}

\end{document}